\documentclass[aps,showpacs,preprintnumbers, superscriptaddress, nofootinbibt,twocolumn]{revtex4}

\usepackage{amsmath}
\usepackage{eurosym}
\usepackage{amssymb}
\usepackage{graphicx}
\usepackage{color}

\def\be{\begin{equation}}
\def\ee{\end{equation}}
\def\bea{\begin{eqnarray}}
\def\eea{\end{eqnarray}}

\begin{document}

\title{The QCD mass gap and quark deconfinement scales as mass bounds in strong gravity  }
\author{Piyabut Burikham}
\email{piyabut@gmail.com}
\affiliation{High Energy Physics Theory Group, Department of Physics, Faculty of Science,
Chulalongkorn University, Phyathai Rd., Bangkok 10330, Thailand}
\author{Tiberiu Harko}
\email{t.harko@ucl.ac.uk}
\affiliation{Department of Physics, Babes-Bolyai University, Kogalniceanu Street,
Cluj-Napoca 400084, Romania}
\affiliation{Department of Mathematics, University College London, Gower Street, London
WC1E 6BT, United Kingdom}
\author{Matthew J. Lake}
\email{matthewj@nu.ac.th}
\affiliation{School of Physics, Sun Yat-Sen University, Guangzhou 510275, China} 
\affiliation{School of Physical and Mathematical Sciences, Nanyang Technological University, 637371 Singapore}
\affiliation{The Institute for Fundamental Study, ``The Tah Poe Academia Institute", \\
Naresuan University, Phitsanulok 65000, Thailand}
\affiliation{Thailand Center of Excellence in Physics, Ministry of Education, Bangkok
10400, Thailand }
\date{\today }

\begin{abstract}
Though not a part of mainstream physics, Salam's theory of strong gravity remains a viable effective model for the description of strong interactions in the gauge singlet sector of QCD, capable of producing particle confinement and asymptotic freedom, but not of reproducing interactions involving $SU(3)$ colour charge. It may therefore be used to explore the stability and confinement of gauge singlet hadrons, though not to describe scattering processes that require colour interactions. It is a two-tensor theory of both strong interactions and gravity, in which the strong tensor field is governed by equations formally identical to the Einstein equations, apart from the coupling parameter, which is of order $1 \ \rm GeV^{-1}$. We revisit the strong gravity theory and investigate the strong gravity field equations in the presence of a mixing term which induces an effective {\it strong cosmological constant}, $\Lambda_{f}$. This introduces a {\it strong de Sitter radius} for strongly interacting fermions, producing a confining bubble, which allows us to identify $\Lambda_{f}$ with the `bag constant' of the MIT bag model, $B \simeq 2 \times 10^{14} \rm gcm^{-3}$. Assuming a static, spherically symmetric geometry, we derive the strong gravity TOV equation, which describes the equilibrium properties of compact hadronic objects. From this, we determine the generalised Buchdahl inequalities for a strong gravity `particle', giving rise to upper and lower bounds on the mass/radius ratio of stable, compact, strongly interacting objects. We show, explicitly, that the existence of the lower mass bound is induced by the presence of $\Lambda_f$, producing a mass gap, and that the upper bound corresponds to a deconfinement phase transition. The physical implications of our results for holographic duality in the context of the AdS/QCD and dS/QCD correspondences are also discussed.

\end{abstract}

\maketitle

\tableofcontents

\section{Introduction} \label{sect1}

One of the most intriguing aspects of short-distance physics is that the strong interactions of hadrons in the infrared (IR) regime exhibit certain features bearing a close resemblance to gravity. For example, string theory -- originally proposed as a theory of strongly interacting hadrons -- can be reinterpreted as a theory of linearised gravity, and the quantum theory of closed bosonic strings naturally includes a candidate graviton \cite{Green:2012pqa}. On the other hand, it is generally assumed that, in low energy physics, the gravitational interaction plays a negligible role. However, it is important to note that the strength of the gravitational interaction increases with energy, the coupling being proportional to $GE^2$, where $E$ is the total energy of the particle (including rest mass) and $G$ is Newton's constant. Hence, gravitational interactions become more and more important at higher energies. In fact, if the particle energy exceeds $E = ec^2/\sqrt{G} \simeq 10^{18}$ GeV, the gravitational interaction is stronger than the electromagnetic interaction and, at energies of the order of $10^{19}$ GeV, it is as strong as the strong nuclear interaction.  The important role of gravitation in fundamental particle physics was also pointed out in \cite{1,2}.

At the quantum level, the gravitational interaction is expected to be mediated by massless spin-2 bosons (gravitons) in a way that is analogous to the mediation of the electromagnetic interaction by massless spin-1 bosons (photons) \cite{3}. Strong support for this idea is provided by the fact that the quantisation of linearised gravity,
obtained by substituting $g_{\mu \nu} = \eta _{\mu \nu} + h_{\mu \nu}$, where $\eta _{\mu \nu}$ is the Minkowski metric and $h_{\mu \nu}$ is an arbitrary perturbation, into the vacuum Einstein equations (neglecting second order quantities), leads to the well-known Pauli-Fierz equations for massless spin-2 particles, $\Box h_{\mu \nu}= 0$, $h_{\mu \nu ,\nu}=0$, $h_{\mu}^{\mu}=0$ \cite{3}.  Thus, from a quantum theoretical point of view, the long range of the gravitational force is a consequence of the masslessness of the mediating particles. This result can be extended to the massive graviton case, and hence it follows that a non-linear self-interacting spin-2 field can also be described by Einstein's field equations  \cite{3a,3b,3c}.

Classically, the (non-vacuum) Einstein field equations, $G_{\mu \nu} = \kappa T_{\mu \nu}$, where $\kappa = 8\pi G/c^4$, relate a covariant geometrical quantity, the Einstein tensor $G_{\mu \nu} = R_{\mu \nu} - (1/2)Rg_{\mu \nu}$, to a covariant physical quantity, the conserved energy-momentum tensor $T_{\mu \nu}$, via the proportionality (coupling) constant $\kappa$ \cite{4}. However, the derivation of the Einstein field equations does not place any restriction whatsoever on the numerical value of the constant $\kappa$. For the canonical gravitational interaction, this must be recovered from the Newtonian limit of the theory \cite{4}.  Noting the existence of strongly interacting, massive, spin-2 meson states (such as the f-meson), and arguing by analogy with the quantisation of linearised Einstein gravity, it was proposed in \cite{sg1,sg1a,sg1b,sg1c} that a short-range `strong gravity' interaction may be responsible for the properties of such elementary particles at a microscopic level.

Thus, a new `metric' tensor $f_{\mu \nu}$, which determines the properties of the strong gravity field, as well as a new strong coupling constant, determined to be of order $1 \ \rm GeV^{-1}$ to ensure consistency with the known physics of strong interactions, were introduced \cite{sg1,sg1a,sg1b,sg1c}. Throughout the rest of this paper, we follow Salam's original (though slightly unconventional) notation, denoting the canonical gravitational and strong gravity coupling constants as
\be \label{coupling_consts}
k_g^2 = 8\pi G/c^4 \, , \quad k_f ^2= 8\pi G_f/c^4 \, ,
\ee
respectively. The dimensionless strong gravity coupling is taken to be of the same order of magnitude as the strong interaction coupling, giving
\be \label{alpha_s}
\alpha_s = \frac{G_fm^2}{\hbar c} \simeq 1 \, .
\ee
This is equivalent to $G_f \simeq 10^{38} G = 6.67 \times 10^{30} \rm cm^3g^{-1}s^{-2}$ for a strong interaction scale of $m \simeq 10^{-19}m_{\rm Pl} \simeq 10^{-24}$ g, where $m_{\rm Pl} = \sqrt{\hbar c/G} \simeq 10^{-5}$ g is the Planck mass.

It is interesting to note that the ratio $G_f/G$ yields `Dirac's number', i.e.
\be \label{Dirac'sNo.}
G_f/G \simeq 10^{38} \, ,
\ee
the same dimensionless quantity that formed the basis of his Large Number Hypothesis (LNH) \cite{Dirac1937,Dirac1938,Dirac1974,Dirac1979}. (See also \cite{MenaMarugan:2001qn,Ray:2007cc,Nassif:2015sdu} for contemporary viewpoints of the LNH.) The possible implications of this, and its relation to more recent results that imply a relation between the cosmological constant and physics at the electroweak scale, {\it {\`a} la} Dirac, are discussed below, and at length in Sec. \ref{sect4}

Hence, in order to describe strong interactions, the following Lagrangian density was proposed \cite{sg1, sg1a,sg1b,sg1c}
\be  \label{eq1}
\mathcal{L} =-\frac{1}{2k_g^2}\sqrt{-g}R(g)-\frac{1}{2k_f^2}\sqrt{-f}R_f(f)+\mathcal{L}_{fg}+\mathcal{L}_m \,
\ee
where $\mathcal{L}_f=R_f(f)$ and $R_f(f)$ is the scalar curvature constructed from $f_{\mu \nu}$ and its derivatives.  The term $\mathcal{L}_{fg}$ describes the interaction between f-mesons and gravitons and $\mathcal{L}_m$ is the Lagrangian for both strongly interacting and non-strongly interacting matter. If one drops the interaction term $\mathcal{L}_{fg}$ and considers that the dominant term in the field equation obtained from (\ref{eq1}) is given by $\mathcal{L}_f$, one obtains an Einstein-type equation for the $f_{\mu \nu}$ field, with the strong coupling constant $k_f^2$ in place of $k_g^2$. This explains the name strong gravity given to the theory. The physical and mathematical properties of the strong gravity model were investigated in \cite{sg2,sg3,sg4,sg5,sg6}. For a comprehensive review of the early results in strong gravity, and the corresponding references,  see \cite{sg7}.

An alternative attempt to describe the physics of strong interactions using a geometric, general relativity-inspired picture is the reformulation of Yang-Mills theory proposed in the so-called `chromogravity' model \cite{cg1,cg2}. In this model, QCD in the IR region is approximated by the exchange of a dressed two-gluon phenomenological field $\mathcal{G}_{\mu \nu}(x) =B_{\mu}^a B_{\nu}^b\eta _{ab}$, where $\eta _{ab}$ is a color-SU(3) metric, and $B_{\mu}^a$ is the dressed gluon field. This model produces colour confinement, explains the successful features of the hadronic string, predicts the spectrum of baryons and mesons with their Regge trajectories, justifies the interacting boson model, and also `predicts' scaling.  The effective Ricci tensor $\mathcal{R}_{\mu \nu}$  constructed from the field $\mathcal{G}_{\mu \nu}$ induces Einsteinian dynamics. Alternative approaches were proposed in \cite{Lun1,Lun2, John} and \cite{Pav}, respectively.  The relations between Yang-Mills fields and Riemannian geometry were also investigated in \cite{Haag1,Haag2}, where it was shown that it is possible to define gauge invariant variables in the Hilbert space of Yang-Mills theories that manifestly implement Gauss' law on physical states.

From a `true' gravitational perspective, the conditions under which upper and/or lower mass bounds exist for different physical systems is of fundamental importance in theoretical general relativity and relativistic astrophysics. A classic result by Buchdahl \cite{Buch} states that, for stable, compact, charge-neutral objects of mass $M$ and radius $R$, the condition
\be \label{Buchdahl}
\frac{2GM}{c^2R}\leq \frac{8}{9} \, ,
\ee
must be satisfied. If the bound (\ref{Buchdahl}) is violated, collapse to a black hole becomes inevitable. In \cite{Rho1} the general relativistic maximum mass of a stable compact astrophysical stellar type object was found to be of the order of  $3.2 \rm M_{\odot}$. This result was obtained  with the use of the principle of causality, requiring that in the dense matter nuclear matter the speed of sound cannot exceed the speed of light, and of Le Chatelier's principle.  The Buchdahl limit (\ref{Buchdahl}) has been extended to include the effects of a nonzero cosmological constant ($\Lambda \neq 0$) \cite{g1}, of the electric charge of the sphere \cite{g2}, and of an anisotropic interior pressure distribution \cite{g3}.  Sharp bounds on the mass/radius ratio for neutral and charged compact objects, with both isotropic and anisotropic pressure distributions, in the presence of $\Lambda \neq 0$, were also obtained in \cite{g4,g5,g6,g7,g8}.

If the existence of an upper mass bound for stellar type structures seems to be a reasonable physical requirement of general relativity, the possible existence of a minimum mass is less obvious. In \cite{m1} it was shown that the presence of a positive cosmological constant implies the existence of a minimum {\it classical} mass and of a minimum density in nature. These results rigorously follow from the generalized Buchdahl inequality in the presence of $\Lambda >0$, given by
\be \label{Buch_Lambda>0}
\sqrt{1-\frac{2GM}{Rc^{2}}-\frac{\Lambda R^2}{3}}\geq \frac{1}{3}-\frac{\Lambda c^{2}}{12\pi G\rho } \, ,
\ee
which implies the existence of a lower bound for the mass/radius ratio or, equivalently, the density of a stable, charge-neutral, gravitating compact object, i.e.
\be \label{min_dens_Lambda>0}
\frac{2GM}{Rc^{2}}\geq \frac{\Lambda }{6}R^2 \iff \rho = \frac{3M}{4\pi R^3} \geq \rho _{\rm min} := \frac{\Lambda c^{2}}{16\pi G} \, .
\ee

Though the derivation of this condition is somewhat involved \cite{m1}, its physical meaning is intuitively obvious. The dark energy density is given by
\be \label{DE_dens}
\rho_{\Lambda} = -p_{\Lambda}/c = \frac{\Lambda c^2}{8\pi G} \, .
\ee
Hence, Eq. (\ref{min_dens_Lambda>0}) simply states that spherical objects with densities significantly lower than the dark energy density have insufficient self-gravity to overcome the effects of dark energy repulsion. For $\rho \lesssim \rho_{\rm min} = \rho_{\Lambda}/2$, the classical radius $R$ becomes unstable and dark energy repulsion blows the object apart. For future convenience, we note that the current experimental value of the vacuum energy density, inferred from observations of high-redshift type 1A supernovae (SN1A), Large Scale Structure (LSS) data from the Sloan Digital Sky Survey (SDSS) and Cosmic Microwave Background (CMB) data from the Planck satellite, is  is $\rho_{\Lambda} = 5.971 \times 10^{-30}\; {\rm g \;cm^{-3}}$ \cite{Betoule:2014frx,Planckresults}. This corresponds to a value of $\Lambda = 1.114 \times 10^{-56} \;\;\rm cm^{-2}$ for the cosmological constant.

The minimum mass/radius bound, in the absence of dark energy (i.e., for $\Lambda = 0$), was extended for the case of charged objects in \cite{m2}, where it was shown that a stable object with charge $Q$ must obey the relation
\be \label{Buch_Q}
\frac{2Mc^{2}}{R} \geq \frac{3}{2}\frac{Q^2}{R^2}\left[1-\frac{Q^2}{36R^2}+\mathcal{O}\left(\frac{Q^2}{R^2}\right)^4\right] \, .
\ee
This result was extended to to include the presence of a cosmological constant ($\Lambda \neq 0$), and the mass/radius ratio of a stable charged object was found to obey the relation
\be \label{Buch_Q_Lambda}
M \geq \frac{3}{4}\frac{Q^2}{Rc^{2}} + \frac{\Lambda R^{3}c^{2}}{12G} \, .
\ee
to leading order in $Q^2/R$. For $\Lambda = 0$, this relation recovers an earlier result due to Bekenstein \cite{Bekenstein:1971ej}, which demonstrated that the expression for the classical radius of a charged particle -- obtained from equating its rest mass with its electrostatic potential energy -- remains rigorously valid in general relativity.

In \cite{B2} and \cite{LakePaterek1, Lake1}, Eq. (\ref{Buch_Q_Lambda}) was combined with minimum length uncertainty relations (MLURs), obtained from gravitational extensions of canonical quantum theory, leading to a minimum mass bound for stable, charged, quantum mechanical and gravitating compact objects of the form
\begin{eqnarray} \label{m_Q}
M \gtrsim M_{\rm Q} = 2^{-1/3}\alpha_{\rm Q}(m_{\rm Pl}^2m_{\rm dS})^{1/3} \, ,
\end{eqnarray}
where
\begin{eqnarray} \label{alpha_Q}
\alpha_{\rm Q} = Q^2/q_{\rm Pl}^2 \, ,
\end{eqnarray}
and $q_{\rm Pl} = \hbar c$ is the Planck charge. Evaluated for $Q = \pm e$, this gives
\begin{eqnarray}  \label{m>me_bound}
M &\gtrsim& 2^{-1/3}\alpha_{e}(m_{\rm Pl}^2m_{\rm dS})^{1/3} = 7.332 \times 10^{-28} \, {\rm g}
\nonumber\\
&\simeq& m_{e} = 9.109 \times 10^{-28} \, {\rm g} \, ,
\end{eqnarray}
where $\alpha_e = e^2/q_{\rm Pl}^2 \simeq 1/137$ is the usual fine structure constant and $m_e$ is the electron mass, which is equivalent to
\begin{eqnarray}  \label{Q^2/me_bound}
\frac{Q^2}{M} &\lesssim& \left(\frac{3\hbar^2G^2c^6}{\Lambda}\right)^{1/6} = 3.147 \times 10^8 \, {\rm Fr^2g^{-1}}
\nonumber\\
&\simeq& \frac{e^2}{m_e} = 2.533 \times 10^{8} \, {\rm Fr^2g^{-1}} \, .
\end{eqnarray}
According to this relation, if the electron were any less massive (with the same charge $e$) or more highly charged (with the same mass $m_e$) a combination of electrostatic and dark energy repulsion would destabilise its Compton wavelength \cite{B2,LakePaterek1, Lake1}. We also note the close similarity between the physical picture of the electron, modelled as an extended charged fluid sphere \cite{Bekenstein:1971ej}, used to derive Eq. (\ref{Q^2/me_bound}) in \cite{B2}, and Dirac's `extensible' model of the electron, proposed in \cite{Dirac:1962iy}. Equation (\ref{Q^2/me_bound}) may also be rewritten as
\begin{eqnarray} \label{Lambda_vs_re}
\Lambda \lesssim \frac{l_{\rm Pl}^4}{r_e^6}=\frac{3\hbar ^2G^2m_e^6c^6}{e^{12}} \simeq 1.366 \times 10^{-56} \;{\rm cm}^{-2} \, ,
\end{eqnarray}
yielding an upper bound for $\Lambda$ which is consistent with the current best fit value inferred from cosmological observations \cite{Planckresults,Betoule:2014frx}. Interestingly, the relation (\ref{Lambda_vs_re}) was previously derived via three different methods (see \cite{Nottale1993,m3,Beck:2008rd}) and, for the mass scale $m=m_e/\alpha_e$, is equivalent to Zel'dovich's estimate of $\Lambda$, based on his reformulation of Dirac's LNH for an asymptotically de Sitter Universe \cite{Zel'dovich:1968zz}. These results suggest the existence of a deep connection between gravity, the presence of a positive cosmological constant, and the stability of fundamental particles, and are discussed further in the context of the strong gravity model, with a `strong cosmological constant' $\Lambda_f$, in Sec. \ref{sect4}.

The particle physics and cosmological implications of the mass scale $M_T=(\hbar^2\sqrt{\Lambda}/G)^{1/3}$, which corresponds to taking $Q^2 \rightarrow q^{2}_{\rm Pl}$ in Eq. (\ref{m_Q}), were considered in \cite{B4}, where, based on an MLUR, it was shown that a black hole with age comparable to the age of the Universe would stop radiating and form a relic state when its mass reaches the dual mass scale $M_T^{'} = m_{\rm Pl}^2/M_T = c(\hbar /G^2\sqrt{\Lambda})^{1/3}$. Moreover, it was shown that a holographic relation exists between the entropy and horizon area of the remnant black hole in generic dimensions.

Though, in the present work, we derive mass bounds in the context of the original strong gravity theory, based on the analogy between general relativity and strong interactions, we note that, in recent years, many theories of modified gravity have been proposed in the literature \cite{mg1,mg2,mg3}. In general, these aim to solve the problems posed by modern observational cosmology without the need to posit the existence of exotic states of matter and energy, i.e. dark matter and dark energy \cite{d1,d2,d3}.  Theoretically, such approaches may also be extended to the physics of strong interactions: if modified gravity theories possess desirable properties from a cosmological perspective, could modified strong gravity theories possess desirable properties from a particle physics point of view?

Though beyond the scope of this paper, we note that upper and lower bounds on the mass/radius ratio of stable compact objects in modified gravity theories were obtained in \cite{B3}, in which modifications of the canonical gravitational dynamics were described by an effective contribution to the matter energy-momentum tensor. As an application of the general formalism developed therein, compact bosonic objects, described by scalar-tensor gravitational theories with self-interacting scalar potentials, and charged compact objects were considered. For Higgs type potentials, it was found that the mass bounds can be expressed in terms of the value of the potential at the surface of the compact object. The general implications of minimum mass bounds for the gravitational stability of fundamental particles and for the existence of holographic duality between bulk and boundary degrees of freedom were also investigated.

It is the goal of this work to investigate the existence of mass bounds in the strong gravity model proposed in \cite{sg1, sg1a,sg1b,sg1c}, and to discuss the relevance of these bounds for hadronic physics and cosmology via the holographic principle. To prove the existence of both minimum and maximum mass bounds, we consider a static, spherically symmetric `geometry' for the strong gravity metric, together with the Einstein gravitational field equations, in which the matter energy-momentum tensor consists of two components: ordinary matter, described thermodynamically by its energy density and anisotropic pressure distribution, and a mixing term. With a specific choice of metric tensor, the coefficients of the contribution from the mixing term take the form of an effective strong cosmological constant, $\Lambda_f$, whose repulsive~(or attractive) force is `felt' only {\it inside} the strongly interacting matter.

After determining the effective Einstein field equations, the Tolman-Oppenheimer-Volkoff (TOV) equation describing the equilibrium properties of the strong gravity system is obtained. With the use of this equation, and adopting some physically reasonable assumptions about the behaviour of the physical and geometrical quantities, we derive the generalized Buchdahl inequality, which is valid at all points inside the compact objects. By evaluating this bound on the surface of the hadronic `particle', we therefore obtain both upper and lower bounds of the mass/radius ratio of the hadrons in the strong gravity model. These bounds depend on the mass parameter (i.e. coupling) in the mixing Lagrangian $\mathcal{L}_{fg}$, as well as of the geometric properties of the hadrons. The physical implications of our results are also discussed.

This paper is organised as follows. The TOV equation for strong gravity is derived in Sec.~\ref{sect2}. The generalised Buchdahl inequality, and the resulting upper and lower bounds on the mass/radius ratio of strongly interacting particles, are derived in Sec.~\ref{sect3}. The strong gravity mass gap, and its implications for holography, are discussed in Sec. \ref{sect4}. In Sec.~\ref{sect5}, we combine the mass bounds obtained in the strong gravity model with MLURs motivated by quantum gravity research, exchanging $G \rightarrow G_f$ and $\Lambda \rightarrow \Lambda_f$ where necessary. Identifying the `strong dark energy density' with the bag constant of the MIT bag model, which is of the order of the nuclear density, $B \simeq 2 \times 10^{14} \ {\rm gcm^{-3}}$, then gives rise to new mass bounds for both neutral and charged strongly interacting particles. Section~\ref{sect6} contains a brief summary and discussion of our main conclusions, of outstanding problems, and of prospects for future work.

\section{Tolman-Oppenheimer-Volkoff Equation in the Strong Gravity Model } \label{sect2}

In the present Section we briefly review the physical basis and mathematical formalism of the strong gravity model, in which it is assumed that a tensor field, obeying an Einstein-type equation, plays a fundamental role in strong interaction physics. In this approach, strong interactions are governed by a set of field equations formally identical to the Einstein equations, apart from the coupling parameter $k_f \simeq 1$ GeV$^{-1}$, which replaces the Newtonian coupling $k_g \simeq 10^{-19}$ GeV$^{-1}$.  Under the assumption of static spherical symmetry we write down the field equations of the model in the presence of an energy-momentum tensor containing an anisotropic fluid term, and derive the generalised Tolman-Oppenheimer-Volkoff (TOV) equation describing the equilibrium properties of stable, compact, hadronic objects. Throughout, we use the sign conventions and the definitions of the geometric tensors given in \cite{4}.

\subsection{Strong gravity} \label{sect2.1}

As stated in the Introduction, the Lagrangian for the interacting strong field metric $f_{\mu \nu}$ and gravitational metric $g_{\mu \nu}$ can be constructed as
\be
\mathcal{L} =-\frac{1}{2k_g^2}\sqrt{-g}R(g)-\frac{1}{2k_f^2}\sqrt{-f}R_f(f)+\mathcal{L}_{fg}+\mathcal{L}_m \,
\ee
where the corresponding `volume element' $\sqrt{-f}$ is defined via $f = {\rm det}f_{\mu \nu}$.  Here, the first term represents the standard general relativistic Lagrangian for the gravitational field, while the second is its strong interaction analogue, obtained by replacing $k_g$ by $k_f$ and $g_{\mu \nu}$ by $f_{\mu \nu}$. To give the `elementary' particles mass, as well as their weak gravitational interaction, a mixing term between the strong and weak gravitational fields, $\mathcal{L}_{fg}$, is needed. $\mathcal{L}_{m}$ represents the matter Lagrangian for both strongly interacting matter and non-strongly interacting matter, where it is assumed that the latter contains terms in $g_{\mu\nu}$ and its derivative {\it only}, whereas the former may depend (generically) on both $g_{\mu\nu}$ and $f_{\mu\nu}$. Hence, although the strong gravity metric minimally couples to {\it all} forms of matter (see below), strongly interacting particles and non-strongly interacting particles `feel' the curvature of the strong metric differently.  A simple covariant mixing term was proposed in \cite{sg3} and is given by
\bea
\mathcal{L}_{fg} &=& \frac{\mathcal{M}^2}{4k_f^2}\sqrt{-g}\left(f^{\mu \nu}-g^{\mu \nu}\right)\left(f^{\kappa \lambda}-g^{\kappa \lambda }\right)
\nonumber\\
&\times& \left(g_{\kappa \lambda }g_{\lambda \nu}-g_{\mu \nu}g_{\kappa \lambda}\right) \, ,
\eea
where $\mathcal{M}$ is a constant with the dimension of mass.  For later use, the full dimension of the mass mixing parameter is given by $\mathcal{M}^{2}\to \mathcal{M}^{2}c^{2}/\hbar^{2}$, the inverse Compton wavelength squared.

In the limit in which the gravitational field may be ignored, $g_{\mu \nu} \rightarrow \eta _{\mu \nu}$, the field equations of the strong gravity theory can be written as
\be
R_{\mu \nu}(f)-\frac{1}{2}f_{\mu \nu}R(f)=k_f^2T_{\mu \nu}^{(s)} \, ,
\ee
where
\be
k_f^2T_{\mu \nu}^{(s)}=-\frac{1}{2}\mathcal{M}^2\left(f^{\kappa \lambda}-\eta ^{\kappa \lambda }\right)\left(\eta _{\kappa \nu }\eta_{\lambda \nu}-\eta_{\mu \nu}\eta_{\kappa \lambda}\right)\frac{\sqrt{-\eta}}{\sqrt{-f}}.
\ee
In the following, we will consider the effect of the strong gravity interaction for a static sphere filled with strongly interacting matter fluid. In spherical polar coordinates $\left\{t,r,\theta,\phi \right\}$ the line element with respect to the strong gravity metric $f_{\mu\nu}$ is assumed to be of the form
\be
dq^{2} = f_{\mu\nu}dx^{\mu}dx^{\nu} = e^{\nu(r)}c^2dt^{2} - e^{\lambda(r)}dr^{2} - \Sigma(r) d\Omega_{2} \, ,
\ee
where $d\Omega_{2}=d\theta^{2}+\sin^{2}\theta d\phi^{2}$ is the line-element for the unit $2$-sphere and $\nu(r)$, $\lambda(r)$ and $\Sigma(r)$ are arbitrary functions of the radial coordinate. Furthermore, we assume that the fluid can be described by the standard energy-momentum tensor
\be
T^{\mu}_{\nu} = \text{diag}(\rho c^2,-P_{r},-P_{\perp},-P_{\perp}) \, ,
\ee
where $\rho c^2$ is the fluid energy density, $P_r$ is the radial pressure and $P_{\perp}$ denotes the tangential pressure.

Ignoring the weak gravitational interaction, the field equations for the strong gravity field coupled to the matter fluid are, therefore,
\be
R^{\mu}_{\nu}(f)-\frac{1}{2}f^{\mu}_{\nu}R(f)=k_f^{2}(T^{\mu (s)}_{\nu}+T^{\mu}_{\nu}) \, ,
\ee
where the raising and lowering of spacetime-like indices is performed via the tensors $f^{\nu\lambda}$ and $f_{\mu\nu}$, respectively, satisfying the condition
\be
f_{\mu\nu}f^{\nu\lambda}=\delta^{\lambda}_{\mu} \, .
\ee
{\it Note the universality of coupling between strong gravity and any form of matter energy-momentum tensor.}

\subsection{Strong gravity field equations} \label{sect2.2}

The stress tensor of the mixing term between massive and massless gravitons can be computed straightforwardly, giving
\begin{eqnarray}
k^{2}_{f}T^{t(s)}_{t}&=&-\frac{\mathcal{M}^{2}}{2}\frac{r^{2}e^{-(\nu+\lambda)/2}}{\Sigma}\left( 3-\frac{2r^{2}}{\Sigma}-e^{-\lambda}\right)e^{-\nu}, \\
k^{2}_{f}T^{r(s)}_{r}&=&-\frac{\mathcal{M}^{2}}{2}\frac{r^{2}e^{-(\nu+\lambda)/2}}{\Sigma}\left( 3-\frac{2r^{2}}{\Sigma}-e^{-\nu}\right)e^{-\lambda}, \\
k^{2}_{f}T^{\theta(s)}_{\theta}&=&-\frac{\mathcal{M}^{2}}{2}\frac{r^{4}e^{-(\nu+\lambda)/2}}{\Sigma^2}\left( 3-\frac{r^{2}}{\Sigma}-(e^{-\nu}+e^{-\lambda})\right), \notag \\
\end{eqnarray}
For simplicity, and also to make $T^{t(s)}_{t}$ interpretable as proper density, we fix the gauge so that \cite{sg3}
\be\label{23}
\Sigma = \frac{2r^{2}}{3} \, , \quad e^{(\nu+\lambda)} = \Delta = {\rm constant} > 0 \, .
\ee

As a result, $T^{r(s)}_{r} = T^{t(s)}_{t}c^2$, giving rise to an equation of state, $P^{(s)}_{r}=-\rho^{(s)}c^2$, which is characteristic for the {\it cosmological constant}.  The particular gauge choice, $\Delta = {\rm const.}$, forces the equation of state of matter to satisfy
\be
P_{r} + \rho c^2 = 0 \, ,
 \ee
since the sum of Eqs.~(\ref{masseq}) and (\ref{nueq}) is identically zero. Such an equation of state is uncommon for ordinary matter, but it is satisfied by a $U(1)$ gauge field in the Coulomb gauge, for example, in the description of a static charged sphere \cite{B2}.  With this choice of gauge, there is an anisotropic Poincar\'{e} stress associated with the $\theta$-component of the energy-momentum tensor, given by
\be
k^{2}_{f}T^{\theta(s)}_{\theta}=\frac{9}{8}\frac{\mathcal{M}^{2}}{\sqrt{\Delta}}\left( \frac{3}{2}-(e^{-\nu}+e^{-\lambda})\right) \, .
\ee
Moreover, from the second of Eqs.~(\ref{23}) it follows that, since $\left(\nu'+\lambda '\right)\Delta =0$ (where a prime denotes differentiation with respect to $r$), the functions $\nu $ and $\lambda$ satisfy the condition
\be
\nu' +\lambda' =0 \, ,
\ee
at all points inside the strongly interacting fluid sphere. The field equations of the strong gravity model may then be written in the form
\begin{eqnarray}
\frac{e^{-\lambda}\lambda'}{r}+\frac{(\frac{3}{2}-e^{-\lambda})}{r^{2}}&=&\frac{3\mathcal{M}^{2}}{4\Delta^{3/2}}+k^{2}_{f}\rho,  \label{masseq} \\
\frac{e^{-\lambda}\nu'}{r}-\frac{(\frac{3}{2}-e^{-\lambda})}{r^{2}}&=&-\frac{3\mathcal{M}^{2}}{4\Delta^{3/2}}+k^{2}_{f}P_{r},  \label{nueq}\\
0=\frac{d}{dr}\left(P_{r}-\frac{3\mathcal{M}^{2}}{4k_f^2\Delta^{3/2}}\right)&+&\frac{\nu'}{2}\left(\rho+P_{r}\right)  \label{coneq} \\
+\frac{2}{r}\Big[ P_{r}-P_{\perp}-\frac{3\mathcal{M}^{2}}{4k_f^2\Delta^{3/2}}&+&\frac{9\mathcal{M}^{2}}{8k_f^2\sqrt{\Delta}}\times \nonumber\\
\left(-\frac{3}{2}+e^{-\nu}+e^{-\lambda}\right)\Big] \, ,
\end{eqnarray}
where the last equation is simply the conservation law, $0=f^{\kappa\mu}\nabla_{\kappa} T_{\mu\nu}$, with respect to the strong gravity metric $f_{\mu\nu}$.

\subsection{The Tolman-Oppenheimer-Volkoff equation in strong gravity} \label{sect2.3}

Eq.~(\ref{masseq}) can be directly integrated to give
\be
e^{-\lambda}=\frac{3}{2}-\frac{\mathcal{M}^{2}}{4\Delta^{3/2}}r^{2}-\frac{k^{2}_{f}M_{0}(r)}{r\Omega_{2}}, \label{lambeq} \,
\ee
where the accumulated mass inside radius $r$ is defined by
\be
M_{0}(r) \equiv \Omega_{2}\int _0^r{\rho r^{2} dr},  \label{masseq1}
\ee
and $\Omega _2=\int d\Omega _2=4\pi$. After substituting $M_{0}$ into Eq.~(\ref{nueq}), we have
\be
\nu' = \left( k^{2}_{f}r^{2}P_{r}-\frac{\mathcal{M}^{2}}{2\Delta^{3/2}}r^2+\frac{k^{2}_{f}M_{0}(r)}{r\Omega_{2}}\right)\frac{e^{\lambda}}{r} \, .  \label{nueq1}
\ee
Substituting $\nu'$ into the conservation law, we obtain the TOV equation for a fluid sphere in the strong gravity model, in the presence of anisotropic stresses, as
\begin{widetext}
\be
\frac{d P_{r}}{dr}=-\frac{(\rho+P_{r})e^{\lambda}}{2r}\left( k^{2}_{f}r^{2}P_{r}-\frac{\mathcal{M}^{2}}{2\Delta^{3/2}}r^2+\frac{k^{2}_{f}M_{0}(r)}{r\Omega_{2}} \right)-\frac{2}{r}\Big[ P_{r}-P_{\perp}-\frac{3\mathcal{M}^{2}}{4k_f^2\sqrt{\Delta}}\Big(\frac{9}{4}+\frac{1-\frac{3}{2}(e^{\nu}+e^{\lambda})}{\Delta}\Big)\Big] \, . \label{TOVeq}
\ee
\end{widetext}
Note that, by taking into account that $\nu +\lambda =0$, Eq.~(\ref{nueq1}) can also be written as
\be
\frac{d}{dr}e^{-\lambda}=\left( k^{2}_{f}r^{2}P_{r}-\frac{\mathcal{M}^{2}}{2\Delta^{3/2}}r^2+\frac{k^{2}_{f}M_{0}(r)}{r\Omega_{2}}\right)\frac{1}{r} \, ,
\ee
which, combined with Eq.~(\ref{lambeq}), gives an alternative definition of the mass of the fluid,
\be
\frac{dM_0}{dr} = -\Omega _2 r^2 P_r \, .
\ee

\section{The Buchdahl inequality and the minimum and maximum mass/radius ratios of stable compact objects in strong gravity} \label{sect3}

In the present section we derive the generalised Buchdahl inequality that constrains the values of the mass and pressure, as well as the geometric quantities of the strong gravity field (i.e. the strong cosmological constant) at an arbitrary point $r$ inside a compact object. From this inequality, both lower and upper bounds on the mass/radius ratio of a static, spherically symmetric object, interacting according to the strong gravity law, can be easily obtained.

\subsection{The Buchdahl inequality} \label{sect3.1}

In order to obtain the generalised Buchdahl inequality for strong gravity, we define the following Buchdahl variables
\begin{eqnarray}
y^{2}\equiv e^{-\lambda}=\frac{3}{2}-2w(r)r^{2}  \, , \quad \zeta \equiv e^{\nu (r)/2}  \, , \quad x\equiv r^{2} \, ,
\end{eqnarray}
where
\be
w(r)=\frac{k^{2}_{f}M_{0}(r)}{2\Omega_{2}r^{3}}+\frac{\mathcal{M}^{2}}{8\Delta^{3/2}} \, .
\ee
From Eqs.~(\ref{lambeq}), (\ref{nueq1}) and (\ref{TOVeq}), we then have
\be
y^{2}\zeta'(x)=\zeta\left( \frac{k^{2}_{f}}{4}P_{\rm eff}+\frac{w(x)}{2} \right) \, ,
\ee
where we have defined
\be
P_{\rm eff}\equiv P_{r}-\frac{3\mathcal{M}^{2}}{4\Delta^{3/2}k^{2}_{f}} \, ,
\ee
which denotes the effective radial pressure.  {\it From here on, a prime indicates differentiation with respect to $x$}, though this convention does {\it not} apply to the notation for the radial coordinate $r^{\prime}$. Further manipulation then leads to
\begin{eqnarray} \label{further_manipulation}
y(y\zeta')' = &&\frac{\zeta w'}{2}+\frac{k^{2}_{f}\zeta (x)}{4x}\Big[ P_{\perp}-P_{r} \notag \\
&&+ \frac{3\mathcal{M}^{2}}{4\sqrt{\Delta}k^{2}_{f}}\Big(\frac{9}{4}+\frac{1-\frac{3}{2}(e^{\nu}+e^{\lambda})}{\Delta}\Big)\Big] \, . \label{prebuch}
\end{eqnarray}

To separate positive and negative terms on the right-hand side of Eq. (\ref{further_manipulation}), we introduce two new quantities, $\gamma $ and $\gamma _{-}$, defined as
\begin{eqnarray}
\frac{\gamma}{r} &\equiv& \frac{k^{2}_{f}\zeta(x)}{x}\left( P_{\perp}-P_{r}+\frac{27\mathcal{M}^{2}}{16\sqrt{\Delta}k^{2}_{f}} \right),  \\
\gamma_{-} & \equiv& \frac{k^{2}_{f}\zeta(x)}{4x}\frac{3\mathcal{M}^{2}}{4\sqrt{\Delta}k^{2}_{f}}\left( \frac{1-\frac{3}{2}(e^{\nu}+e^{\lambda})}{\Delta} \right).
\end{eqnarray}
From Eq.~(\ref{lambeq}), it follows that, since $e^{\lambda}\geq 2/3$ and $e^{\nu}\geq 0$, $\gamma_{-}$ is always negative. On the other hand, $\gamma$ is positive definite for $P_{\perp}>P_{r} - 27\mathcal{M}^2/16\Delta^{3/2}k_f^2$.  For a static charged sphere, $P_{\perp}=-P_{r}=Q^{2}(r)/2r^{4}$~\cite{B2}, where $Q(r)$ is the accumulated charge, this condition is valid as long as the pressure from the mass mixing contribution satisfies
\be
\frac{27\mathcal{M}^{2}}{16\sqrt{\Delta}k^{2}_{f}} > P_{r}-P_{\perp} = -\frac{Q^{2}}{r^{4}}.
\ee
Note that the mixing term could be negative.

Equation (\ref{prebuch}) can be further simplified by defining
\begin{eqnarray}
dz&\equiv&\frac{1}{y}dx=\frac{2r}{y}dr \, ,  \notag \\
\psi &\equiv& \zeta - \eta \, ,  \notag \\
\eta &\equiv& \int^{r}_{0}\left( \int^{r_{1}}_{0}\frac{\gamma(r_{2})}{y(r_{2})}~dr_{2} \right)\frac{r_{1}}{y(r_{1})}~dr_{1} \, ,
\end{eqnarray}
to obtain
\be
\frac{d^{2}\psi(z)}{dz^{2}}=\frac{w'(x)\zeta}{2}+\gamma_{-} \, . \label{bucheq}
\ee
Assuming monotonically decreasing profiles for both the density $\rho(r)$ and $\gamma(r)$, it follows that for all $r>r'$  the conditions
\be
\frac{M_{0}(r')}{r'}>\frac{M_{0}(r)}{r}\left( \frac{r'}{r}\right)^{2} \, , \quad \gamma(r')>\gamma(r) \, ,  \label{condinq}
\ee
must hold at all points inside the compact object. We then immediately obtain $w'(x)<0$ and, thus, it follows that the right-hand side of Eq.~(\ref{bucheq}) is always negative. Using the mean value theorem, we therefore obtain the following inequalities for the first and second derivatives of $\psi$ with respect to $z$,
\be\label{46}
\frac{d^{2}\psi(z)}{dz^{2}}<0 \longleftrightarrow \frac{d\psi}{dz}\leq \frac{\psi(z)-\psi(0)}{z} \, .
\ee
For $\psi(0)=\zeta(0)-\eta(0)>0$, this leads to
\begin{widetext}
\be
\frac{y}{2r}\frac{d\zeta}{dr}-\frac{1}{2}\int \frac{\gamma(r)}{y(r)}~dr < \left(\int \frac{2r}{\sqrt{\frac{3}{2} - 2 w r^{2}}}~dr\right)^{-1}\left[ \zeta-\int^{r}\left(\int^{r_{1}}\frac{\gamma(r_{2})}{y(r_{2})}~dr_{2} \right)\frac{r_{1}}{y(r_{1})}~dr_{1}, \right] \, .  \label{buchin}
\ee
\end{widetext}
Using condition (\ref{condinq}), we find
\begin{eqnarray}
\int^{r}_{0} \frac{\gamma(r')}{y(r')}~dr' &\geq& \gamma(r)\sqrt{\frac{2}{3}}\int^{r}_{0}\frac{dr'}{\sqrt{1-\frac{2 m_{\rm eff}}{r^{3}}r'^{2}}} \notag \\
&\geq&\gamma(r) \left( \frac{r^{3}}{3m_{\rm eff}}\right)^{1/2}\arcsin{\sqrt{\frac{2m_{\rm eff}}{r}}} \, , \notag \\  \label{app1}
\end{eqnarray}
where
\bea
m_{\rm eff}(r) &\equiv& \frac{1}{3}\left(k^{2}_{f}\int _0^r\rho r^{2}~dr + \Lambda_f r^{3}\right)
\nonumber\\
&=& \frac{1}{3}\left(\frac{k_f^2}{\Omega _2}M_0(r)+\Lambda_f r^3\right) \, ,
\eea
and
\be \label{strong_Lambda}
\Lambda_f \equiv \frac{\mathcal{M}^{2}}{4\Delta^{3/2}} \, ,
\ee
is the {\it effective cosmological constant of the strong gravity model}.

Similarly,
\begin{widetext}
\begin{eqnarray}
&&\int^{r}\left( \int^{r_{1}}\frac{\gamma(r_{2})}{y(r_{2})}~dr_{2} \right)\frac{r_{1}}{y(r_{1})}~dr_{1}\geq \gamma(r)\int^{r}_{0}\left( \int^{r_{1}}_{0}\Big[\frac{3}{2}-\frac{3m_{\rm eff}}{r^{3}}r^{2}_{2}\Big]^{-1/2}~dr_{2} \right)\frac{r_1}{y(r_{1})}~dr_{1}  \notag \\
&\geq& \gamma(r)\left( \frac{r^{3}}{3m_{\rm eff}}\right)^{1/2}\int^{r}_{0}dr_{1}r_{1}\left(\frac{3}{2}-\frac{3m_{\rm eff}}{r^{3}}r^{2}_{1} \right)^{-1/2}\arcsin{\sqrt{\frac{2m_{\rm eff}}{r^{3}}}r_{1}}=\gamma(r)\left( \frac{r^{3}}{3m_{\rm eff}}\right)^{3/2}\Big[ \sqrt{\frac{3m_{\rm eff}}{r}}-y \arcsin{\sqrt{\frac{2m_{\rm eff}}{r}}} \Big] \, .  \notag \\  \label{app2}
\end{eqnarray}
\end{widetext}
Finally, we substitute Eqs.~(\ref{app1}), (\ref{app2}) and (\ref{nueq1}) into the inequality (\ref{buchin}) and divide by $\zeta$ to obtain the Buchdahl inequality in strong gravity as
\begin{widetext}
\be
\Bigg\{\left(1-\frac{2m_{\rm eff}}{r}\right)^{-1/2}-1\Bigg\}\frac{\left( k^{2}_{f}r^{3}P_{\rm eff}+3m_{\rm eff} \right)}{3r^{3}} < \frac{2m_{\rm eff}}{r^{3}}+\frac{2k^{2}_{f}\mathcal{D}}{3}\left( \frac{\arcsin{\sqrt{\frac{2m_{\rm eff}}{r}}}}{\sqrt{\frac{2m_{\rm eff}}{r}}}-1 \right) \, ,  \label{buchinq0}
\ee
\end{widetext}
where
\be \label{D}
\mathcal{D} \equiv P_{\perp}-P_{r}+\frac{27}{16}\frac{\mathcal{M}^{2}}{\sqrt{\Delta}k_f^2} \, .
\ee
Equation~(\ref{buchinq0}) is valid for all $r$ inside the strong gravity particle. Moreover, its validity  does not depend on the sign of $\mathcal{D}$.

\subsection{The minimum and maximum mass/radius ratios of hadrons} \label{sect3.2}

\subsubsection{The upper bound on the mass/radius ratio of hadrons} \label{sect3.2.1}

As a simple application of the Buchdahl inequality in strong gravity (\ref{buchinq0}), we consider the {\it quasi-isotropic} limit $\mathcal{D}=0$, corresponding to the condition
\be
P_{\perp}(r)=P_{r}(r)-\frac{27}{16}\frac{\mathcal{M}^{2}}{\sqrt{\Delta}k_f^2} \, .
\ee
Moreover, we assume that {\it the effective pressure also vanishes at the surface of the massive particle}, so that
\bea \label{conds}
P_{\rm eff}=0 \, , \quad P_{r}(R)=\frac{3\mathcal{M}^{2}}{4\Delta^{3/2}k^{2}_{f}} \, ,
\nonumber\\
P_{\perp}(R)=\frac{3\mathcal{M}^{2}}{4\sqrt{\Delta}k^{2}_{f}}\left(\frac{1}{\Delta}-\frac{9}{4}\right) \, .
\eea
By evaluating Eq. (\ref{buchinq0}) at the surface of the hadron $r=R$, using the conditions (\ref{conds}), we obtain
\begin{equation}
\frac{1}{\sqrt{1-\frac{2M_{\rm eff}}{R}}}\leq 2\left [1-\left (1-\frac{2M_{{\rm eff}}}{R}\right )^{\frac{1}{2}}\right ]^{-1} \, ,
\end{equation}
where we have denoted $M_{{\rm eff}}=m_{{\rm eff}}(R)$, leading to the well-known result $2M_{{\rm eff}}/R\leq 8/9$ \cite{Buch}. This shows that, in strong gravity, the maximum possible mass/radius ratio for hadrons should be constrained (at least approximately) by a Buchdahl-type relation. Written in a dimensional form, the Buchdahl inequality for strongly interacting particles of mass $M_{\rm eff}$ and radius $R$ can be written as
\be
\frac{2 G_f}{c^2}\frac{M_{\rm eff}}{R}\leq \frac{8}{9}=0.88^{.} \, .
\ee
For $G_f=6.67\times 10^{30} \ \rm cm^3g^{-1}s^{-2}$, this relation is obviously satisfied in the case of proton, with mass $m_p=1.672 \times 10^{-24}$ g and {\it classical} radius $r_p = 0.875\times 10^{-13}$ cm, such that $2G_fm_p/c^2r_p=0.288$. Interestingly, a particle radius around 3.2 times smaller than $r_p$ would make the proton unstable from the point of view of strong interactions.

Next we consider the case $P_{\rm eff}\neq 0$. In the quasi-isotropic limit $\mathcal{D}=0$, Eq.~(\ref{buchinq0}) gives the upper mass-radius bound
\be
\frac{2M_{\rm eff}}{R}\leq 1-\frac{1}{9}\left[\frac{1+k_f^2P_{\rm eff}(R)/3\left<\rho _{\rm eff}\right>}{1+k_f^2P_{\rm eff}(R)/9\left<\rho _{\rm eff}\right>}\right]^2 \, ,
\ee
where we have defined the mean density of the compact object as $\left<\rho _{\rm eff}\right>=M_{\rm eff}/R^3$. {\it We assume that the matter radial pressure $P_r$ vanishes at the surface of the strong gravity particle}, and thus we obtain for the surface effective pressure the expression
\be\label{76a}
P_{\rm eff}(R)=-\frac{3\mathcal{M}^2}{4k_f^2\Delta ^{3/2}}=-3\frac{\Lambda _f}{k_f^2} \, .
\ee
Taking into account that
\be\label{77}
M_{\rm eff}=\frac{1}{3}\left[\frac{k_f^2}{\Omega _2}M_0(R)+\Lambda _fR^3\right] \, ,
\ee
\be
\left<\rho _{\rm eff}\right>=\frac{1}{3}\frac{k_f^2}{\Omega _2}\left<\rho _0\right>+\frac{1}{3}\Lambda _f \, ,
\ee
\be
\frac{k_f^2P_{\rm eff}}{\left<\rho _{\rm eff}\right>}=-3\frac{\Lambda _f}{\left<\rho _{\rm eff}\right>}=-\frac{9}{1+k_f^2\left<\rho _0\right>/\Omega _2\Lambda _f} \, ,
\ee
where we have defined the mean fluid density as $\left< \rho _0\right>=M_0/R^3$, with $M_0\equiv M_0(R)$,
we obtain the following upper limit for the ordinary matter mass/radius ratio of a stable compact object in strong gravity,
\begin{equation}\label{24a}
\frac{k_f^2M_0}{\Omega _2R}\leq \frac{3}{2}\left (1-\frac{2}{3}\Lambda_f R^{2}\right )\left [1-\frac{1}{9}\frac{\left (1-2\Omega _2\Lambda_f /k_f^2\langle\rho_0\rangle \right )^{2}}{1-\frac{2 }{3}\Lambda_f R^{2}}\right ] \, .
\end{equation}
Next, we consider the anisotropic case with $\mathcal{D} \neq 0$. We define the function $f(M_{{\rm eff}},R,\Lambda_f,\mathcal{D})$ as
\begin{equation}
f\left (M_{{\rm eff}},R,\Lambda_f,\mathcal{D}\right)=\frac{k^{2}_{f}\mathcal{D} \left (R\right)}{3 \langle \rho _{\rm eff}\rangle }\left \{\frac{\arcsin\left [\sqrt{\frac{2M_{{\rm eff}}}{R}}\right ]}{\sqrt{\frac{2M_{{\rm eff}}}{R}}} -1\right\} \, .
\end{equation}
{\it Assuming again that the effective pressure vanishes at the surface of the compact object}, $P_{\rm eff}\equiv 0$, Eq.~(\ref{buchinq0}) leads to the following general restriction on the mass/radius ratio for a spherical hadronic fluid,
\be
\frac{2M_{\rm eff}}{R}\leq 1-\frac{1}{\left[1+2(1+f)\right]^2}=1-\frac{1}{9}\frac{1}{\left(1+2f/3\right)^2}.
\ee
By taking into account the definition of the total effective mass as given by Eq.~(\ref{77}), we immediately find
\bea \label{24}
\frac{k_f^2M_0}{\Omega _2R}&\leq &\frac{3}{2}\left (1-\frac{2 }{3}\Lambda_f R^{2}\right )\times \nonumber\\
&&\left [1-\frac{1}{9}\frac{1}{\left (1-\frac{2 }{3}\Lambda_f R^{2}\right )\left (1+\frac{2f}{3}\right)^{2}}\right] \, .
\eea

With the use of the Taylor series expansion of the function $\arcsin x/x-1$,
\be \label{expansion}
\frac{\arcsin x}{x}-1=\frac{x^2}{6}+\frac{3 x^4}{40}+O\left(x^6\right) \, ,
\ee
for small values of the argument, we can approximate the function $f\left (M_{{\rm eff}},R,\Lambda_f,\mathcal{D}\right)$ as
\be
f\left (M_{{\rm eff}},R,\Lambda_f,\mathcal{D}\right ) \simeq \frac{1}{9}k_f^2\mathcal{D}(R)R^2.
\ee
\\
Then the maximum mass bound for compact objects in strong gravity {\it with vanishing surface effective pressure} can  be reformulated as
\bea
\hspace{-0.8cm}&&\frac{k_f^2M_0}{\Omega _2R}\leq \frac{3}{2}\left (1-\frac{2 }{3}\Lambda_f R^{2}\right )\times \nonumber\\
\hspace{-0.8cm}&&\left [1-\frac{1}{9}\frac{1}{\left (1-\frac{2 }{3}\Lambda_f R^{2}\right )\left (1+2k_f^2\mathcal{D}(R)R^2/27\right)^{2}}\right] \, .
\eea

Finally, we consider the case of strong gravity compact objects {\it with vanishing surface radial pressure}, i.e. with $P_r(R)=0$, and $P_{\rm eff}\neq 0$, given by Eq.~(\ref{76a}). In this case, we obtain for the maximum mass/radius bounds the expressions
\be
\frac{2M_{\rm eff}}{R}\leq 1-\frac{1}{9}\left[\frac{1+k_f^2P_{\rm eff}(R)/3\left<\rho _{\rm eff}\right>}{1+2f/3+k_f^2P_{\rm eff}(R)/9\left<\rho _{\rm eff}\right>}\right]^2,
\ee
and
\bea
&&\hspace{-0.5cm}\frac{k_f^2M_0}{\Omega _2R }\leq \frac{3}{2}\left (1-\frac{2}{3}\Lambda_f R^{2}\right )\times \nonumber\\
&&\hspace{-0.5cm}\left \{1-\frac{1}{9}\frac{\left (1-2\Omega _2\Lambda_f /k_f^2\langle\rho_0\rangle \right )^{2}}{\left(1-\frac{2 }{3}\Lambda_f R^{2}\right)\left[1+2f\left(1+\Omega _2\Lambda _f/k_f^2\left<\rho _{0}\right>\right)/3\right]^2}\right \}, \nonumber\\
\eea
respectively.

The maximum mass/radius bound for compact objects is generally obtained in the constant density regime, with $\rho _{\rm eff} \simeq \rho _0={\rm constant}$. Hence the total mass of the compact object can be approximated as $M_0=4\pi \rho _0 R^3/3$. Therefore, in the mass/radius ratio bounds obtained above, we can approximate the effective mean density as $\left< \rho _{\rm eff}\right > \simeq M_0/R^3 \simeq \rho _0 = {\rm constant}$, a relation that is satisfied by the maximum mass objects with a very good approximation. Therefore, in all the above results, the ratio $\Lambda_f /\left< \rho _{\rm eff}\right >$ can then be approximated as a constant. Hence, it follows that, generally, the right-hand sides of the upper bounds on the mass/radius ratio can be regarded as independent of the masses of the compact objects.

\subsubsection{The lower bound on the mass/radius ratio of hadrons} \label{sect3.2.2}

On the vacuum boundary of the anisotropic fluid distribution, $r=R$, Eq.~(\ref{buchinq0}) takes the general form
\begin{widetext}
\begin{equation}
\sqrt{1-\frac{2M_{\mathrm{eff}}}{R}}\geq \frac{k_{f}^{2}P_{\mathrm{eff}}/3+M_{\mathrm{eff}}/R^{3}}{3M_{\mathrm{eff}}/R^{3}+k_{f}^{2}P_{\mathrm{eff}}/3+2k^{2}_{f}\mathcal{D}(R)\left\{ \arcsin \left[ \sqrt{\frac{2M_{\mathrm{eff}}}{R}}\right] /%
\sqrt{\frac{2M_{\mathrm{eff}}}{R}}-1\right\} /3} \, ,  \label{an1}
\end{equation}
\end{widetext}
where we have assumed that $k_{f}^{2}P_{\mathrm{eff}}/3+M_{\mathrm{eff}}/R^{3}>0$. In addition, we assume that the surface radial pressure $P_{r}$ is negligibly small on the hadron's surface, so that $P_{\mathrm{eff}}(R)=-3\Lambda_f/k^{2}_{f}$, and therefore $P_{\mathrm{eff}}(R)<0$. Again using the fact that, for small values of the argument, the function $\arcsin x/x-1$ can be approximated using Eq. (\ref{expansion}), and performing the replacement $P_{\text{eff}}\to -P_{\text{eff}}$, Eq.~(\ref{an1}) can be written as
\begin{equation}
\sqrt{1-\frac{2M_{\mathrm{eff}}}{R}}\geq \frac{M_{\mathrm{eff}}/R-k_{f}^{2}P_{\mathrm{eff}}R^{2}/3}{\left[ 3+(2k^{2}_{f}/9)\mathcal{D}(R)R^{2}\right] M_{\mathrm{eff}}/R-k_{f}^{2}P_{\mathrm{eff}}R^{2}/3} .  \label{an2}
\end{equation}

By introducing a new variable $v$, defined as
\begin{equation}
v=\frac{M_{\mathrm{eff}}}{R} \, ,
\end{equation}
Eq.~(\ref{an2}) takes the form
\begin{equation}
\sqrt{1-2v}\geq \frac{v-p}{qv-p} \, ,  \label{an3}
\end{equation}
where we have denoted
\be
p = \frac{1}{3}k_{f}^{2}P_{\mathrm{eff}}R^{2} = \Lambda_f R^2=\frac{\mathcal{M}^2}{4\Delta ^{3/2}}R^2 \, ,
\ee
and
\be
q = 3 + \frac{2k^{2}_{f}}{9}\mathcal{D}(R)R^{2} \, ,
\ee
respectively. Then, by squaring Eq. (\ref{an3}), we can reformulate the corresponding inequality as
\begin{equation}
v\left[ 2q^{2}v^{2}-\left( q^{2}+4pq-1\right) v+2p\left( p+q-1\right) \right] \leq 0 \, ,
\end{equation}
or, equivalently,
\begin{equation}
v\left( v-v_{1}\right) \left( v-v_{2}\right) \leq 0 \, ,  \label{cond2}
\end{equation}
where we have denoted
\begin{equation}
v_{1}=\frac{q^{2}+4pq-1-\left( 1-q\right) \sqrt{(1+q)^{2}-8pq}}{4q^{2}} \, ,
\end{equation}
and
\begin{equation}
v_{2}=\frac{q^{2}+4pq-1+\left( 1-q\right) \sqrt{(1+q)^{2}-8pq}}{4q^{2}} \, .
\end{equation}

In the following analysis, we keep only the first order terms in both $p$ (depending on $\Lambda_f$) and $q$ (depending on $\mathcal{D}$) in the expressions involving square roots. Since $v\geq 0$, Eq.~(\ref{cond2}) is satisfied if $v\leq v_{1}$ and $v\geq v_{2}$, or $v\geq v_{1}$ and $v\leq v_{2}$. However, one can easily check that the condition $v\geq v_{1}$ contradicts the upper bound on the mass/radius given by Eq.~(\ref{24}). Therefore, it follows that Eq.~(\ref{cond2}) is identically satisfied if and only if, for all values of the physical parameters determining the total mass of the hadronic particle, the condition $v\geq v_{2}$ holds. This result is equivalent to the existence of a minimum bound for the
mass/radius ratio of particles in strong gravity, which is given by
\begin{equation}
v\geq \frac{2p}{1+q} \, .
\end{equation}
By explicitly substituting the expressions for $p$, $q$ and $v$, as defined above, we then obtain the following (alternative) form of lower bound for the mass/radius ratio of hadronic particles in strong gravity,
\begin{equation}
\frac{k_f^2M_{{\rm eff}}}{\Omega _2R} \geq \frac{1}{2}\Lambda_f R^{2}\frac{1}{1+ (k^{2}_{f}/18)\mathcal{D}(R)R^2} \, .  \label{min}
\end{equation}
By taking into account the explicit expression for the effective mass of the hadron, Eq. (\ref{min}) may be rewritten as
\be \label{76}
\frac{k_f^2M_0}{\Omega _2R}\geq \frac{1}{2}\Lambda_f R^2\frac{1-(k^{2}_{f}/9)\mathcal{D}(R)R^2}{1+(k^{2}_{f}/18)\mathcal{D}(R)R^2} \, .
\ee

Hence, we see that the presence of the anisotropic matter distribution weakens the lower bound on the hadron mass. Nonetheless, in the strong gravity theory, there still exists an absolute minimum mass for hadrons in nature. If the surface anisotropy, described by the coefficient $\mathcal{D}$, can be neglected, the existence of a minimum mass is determined by the presence of the effective strong gravity cosmological constant {\it only}. This is constructed from the mass parameter of the model, $\mathcal{M}$, and the arbitrary constant $\Delta$, which fixes the value of the metric tensor coefficients inside the hadrons. For $\Lambda_f \equiv 0$, the minimum mass bound simply reduces to the positivity requirement for the bare mass, $M_0\geq 0$.

By taking into account the explicit expression for the strong gravity cosmological constant and by assuming that $(k^{2}_{f}/9)\mathcal{D}(R)R^2 \ll 1$, i.e. assuming that the pressure anisotropy vanishes at the vacuum boundary of the hadron, the minimum mass/radius ratio of hadronic particles, given by Eq.~(\ref{76}), can be written as
\be
\frac{G_f M_0}{c^2R}\geq \frac{\mathcal{M}^2}{16\Delta ^{3/2}}R^2 \, .
\ee
This equation imposes strong (pun intended) constraints on the mass parameter $\mathcal{M}$ which controls the strength of the mixing between the $f_{\mu\nu}$ and $g_{\mu\nu}$ fields, i.e.
\be
\mathcal{M}^2 \leq \frac{16G_f}{c^2}\Delta^{3/2}\rho _{{\rm min}} \, ,
\ee
where $\rho _{{\rm min}}=M_0^{({\rm min})}/R^3$ is the density corresponding to the minimum mass hadron.

\subsection{The energy localization problem in strong gravity} \label{sect3.3}

An important issue in general relativity is the problem of energy localization. Tentatively, we assume that the total effective energy in strong gravity can be described in a similar way as in canonical Einstein gravity. This assumption allows us to derive explicit limits on the total energy of compact hadronic objects. Hence, we define the total energy inside an equipotential surface $S$, which includes the contribution from the strong tensor field $f_{\mu\nu}$, by analogy with general relativity, as \cite{32,33}
\begin{equation}
E_f = E_{M}+E_{F}=\frac{1}{8\pi }\xi _{s}\int_{S}\left[ K\right] dS \, , \label{47}
\end{equation}
where the vector $\xi ^{i}$ is a Killing field of time translation, $\xi_{s}$ denotes its value at $S$, and $\left[K\right] $ is the jump across the shell of the trace of the extrinsic curvature of $S$, assumed to be embedded in the 2-space $t=\mathrm{constant}$. $E_{M}=\int_{S}T_{i}^{k}\xi ^{i}\sqrt{-g}dS_{k}$ and $E_{F}$ are the energies of the ordinary matter and of the strong gravitational field in the $f_{\mu \nu}$ metric, respectively. This definition of the total energy is manifestly
coordinate invariant. In the case of static spherical symmetry, for both the $g_{\mu \nu}$ and $f_{\mu \nu}$ fields, we obtain the total energy for the hadron, from Eq.~(\ref{47}), as \cite{33}
\begin{equation}
E_f=-re^{\nu /2}\left[ e^{-\lambda /2}\right]_S \, ,
\end{equation}
where, as usual, $[\;\;]_S$ denotes the jump across the surface $S$. We also make the fundamental assumption that the metric outside the strong gravity system is of de Sitter type, under the replacement $\Lambda_f \rightarrow \Lambda$.

Next, for the sake of convenience, we rescale the metric tensor component so that $(2/3)e^{-\lambda}\rightarrow e^{-\lambda}$. Eq.~(\ref{lambeq}), which may be expressed in terms of the effective mass of the strong gravity object $M_{\rm eff}$, can then be written as
\be
e^{-\lambda}=\left(1-\frac{2M_{\rm eff}}{R}\right) \, .
\ee
Such a rescaling is always possible, and is performed explicitly via coordinate transformation $r^2 \rightarrow r'^{2}=2r^{2}/3$. Note that the mass per unit radial distance transforms as $M(r')/r' = 2M(r)/3r$. For convenience, we will also redefine $\Lambda_{f}\to \Lambda_{f}/3$.

Then, by taking into account the relation $\nu +\lambda =0$, as well as the definition of the strong gravity cosmological constant, the total energy of a compact self-gravitating object may be written as
\be
E_f=R\left(1-\frac{2M_{\rm eff}}{R}\right)^{1/2}\left[1-\left(1-\frac{2M_{\rm eff}}{R}\right)^{1/2}\right],
\ee
or, equivalently,
\bea\label{toten}
E_f&=&R\left (1-\frac{k_f^2M_0}{\Omega _2R}-\frac{\Lambda_f R^2}{3}\right)^{1/2}
\nonumber\\
&\times& \left [1-\left (1-\frac{k_f^2M_0}{\Omega _2R}-\frac{\Lambda_f R^2}{3}\right )^{1/2}\right ] .
\eea

With the use of Eq.~(\ref{buchinq0}), we find the following upper limit for the total energy of the compact hadronic object in strong gravity,
\bea\label{109a}
E_f &\leq& 2R\frac{\left(1+f\right)\left(1-2M_{\rm eff}/R\right)}{1+k_f^2P_{\rm eff}(R)/3\left<\rho _{\rm eff}\right>}.
\eea
In the case of a vanishing strong cosmological constant, $\Lambda_f \rightarrow 0$, and also assuming that the matter pressure is zero at the vacuum boundary of the object, from Eq.~(\ref{109a}) we obtain the strong gravity equivalent of the standard upper energy bound \cite{32,33}
\begin{equation}
E_f \leq 2R\left (1-\frac{k_f^2M_0}{\Omega _2R}\right ) \, .
\end{equation}
For a quasi-isotropic matter distribution with $\mathcal{D} =0$, with the assumption of vanishing radial pressure $P_r(R)=0$, we obtain
\begin{equation}
E_f \leq \frac{2R}{1-\Lambda_f/3\langle \rho _{\rm eff}\rangle }\left (1-\frac{k_f^2M_0}{\Omega _2R}-\frac{1}{3}\Lambda_fR^2\right ).
\end{equation}

Eq.~(\ref{toten}) also allows to obtain a mass/radius relation for hadronic objects in strong gravity  by requiring that the particle is in its minimum energy state, corresponding to $\partial E/\partial R=0$. This gives the following mass/radius relation, as a function of the strong cosmological constant, which is valid for $2\Lambda_f R^2 >1$,
\bea
\frac{k_f^2M_0^{\pm}}{\Omega _2R}&=&2\Lambda_{f} R^{2} \Bigg(\frac{4}{3} - \Lambda_{f} R^{2}\Bigg)\pm 2 R\sqrt{\Lambda_{f} \left(\Lambda_{f}R^2-1\right)^{3}}. \notag \\
\eea
The two values of the mass differ by a quantity
\be
\Delta M_0=M_0^{+}-M_0^{-}=  \frac{4 R^{2} \Omega _2 \sqrt{\Lambda _f \left(   \Lambda _f R^2-1\right)^3}}{k_f^2}.
\ee

\section{The mass gap in strong gravity} \label{sect4}

In Section \ref{sect3.2.1}, we have considered the upper bound on the mass/radius ratio in the situation where $P_{\rm eff}(R)=0$ and $\mathcal{D}=0, P_{\rm eff}\neq 0$. In Section \ref{sect3.2.2}, the lower bound is derived when we set $P_{r}(R)=0$ in order to highlight the effect of anisotropic parameter $\mathcal{D}$ on the bound. In this section, we will consider the most generic case, without making any assumptions about the value of pressure at the surface of the object, and will simply define $P_{\rm eff}(R)\equiv P_{\rm eff}$.

In order to simplify our formalism, we define the additional dimensionless quantities
\begin{eqnarray}
u \equiv \frac{k^{2}_{f}M_{0}(R)}{\Omega_{2}R} \, , \quad b\equiv \Lambda_f R^{2} \, ,
\notag \\
a\equiv \frac{k^{2}_{f}P_{\rm eff}R^{2}}{2} \, , \quad F \equiv \frac{k^{2}_{f}\mathcal{D}R^{2}}{4} \, ,
\end{eqnarray}
which allows us to express the inequality (\ref{buchinq0}) at $r=R$ as
\be
\sqrt{1-\frac{2(u+b)}{3}} > \frac{(u+b+2a)^{2}}{[\left( 3+\frac{8F}{9}\right)(u+b)+2a]^{2}} \, .
\ee
This may be written in the following, explicitly quadratic form,
\begin{eqnarray}
0 &>&(u+b)^{2}+\mathcal{B}(u+b)+\mathcal{C} \, ,  \label{quadeq}
\end{eqnarray}
where
\begin{eqnarray}
\mathcal{B} &=&\left( \frac{3}{2 (3+8F/9)^{2}}+\frac{4a}{(3+8F/9)}-\frac{3}{2} \right) \, , \notag \\
\mathcal{C} &=&\frac{2a}{(3+8F/9)^{2}}\left(2a-6-\frac{8F}{3}\right) \, .
\end{eqnarray}
The inequality can be satisfied if
\be
u_{1}<u<u_{2} \, ,
\ee
for
\begin{eqnarray}
u_{1}&=& -\frac{\mathcal{B}}{2}\left(1-\sqrt{1-\frac{4\mathcal{C}}{\mathcal{B}^{2}}}\right) - b \, , \notag \\
u_{2}&=& -\frac{\mathcal{B}}{2}\left(1+\sqrt{1-\frac{4\mathcal{C}}{\mathcal{B}^{2}}}\right) - b \, ,
\end{eqnarray}
for $\mathcal{B} < 0$. For $\mathcal{B}>0$, the mass bounds cease to exist.  The condition $\mathcal{B}<0$ leads to the following constraint on $F$,
\begin{eqnarray}
F &>& \frac{3}{8}(4a-9+\sqrt{9+16a^{2}}) \, , \quad {\rm for}\quad a>0 \, , \notag  \\
F &>& \frac{3}{8}(4a-9-\sqrt{9+16a^{2}}) \, , \quad {\rm for}\quad a<0 \, .  \label{Fcond}
\end{eqnarray}

Another condition for the existence of mass bounds is $\mathcal{B}^{2}>4\mathcal{C}$, which is trivially satisfied for $a>0$. However, for $a<0$, $\mathcal{B}^{2}>4\mathcal{C}$ requires
\be
F > \frac{3}{2}(\sqrt{4a^{2}+3a}-3-2a) \, .
\ee
To simplify these results, we consider the case when $4\mathcal{C}/\mathcal{B}^{2} \ll 1$ and also $a,F \ll 1$, to obtain
\begin{eqnarray}
u_{1}&\simeq& -\frac{\mathcal{C}}{\mathcal{B}}-b\simeq -a-b+\frac{2a}{3}\left(\frac{F}{3}-a\right) \, ,  \\
u_{2}&\simeq& -\mathcal{B}+\frac{\mathcal{C}}{\mathcal{B}}-b \simeq \frac{4}{3}-b-\frac{a}{3}+\frac{8F}{81} \, .
\end{eqnarray}
Under these conditions, the contribution from the anisotropic stress, $F=k^{2}_{f}\mathcal{D}R^{2}/4$, only appears at the next-to-leading order for the lower bound. Finally, the mass bounds for small $a$, $b$ and $F$ are
\begin{eqnarray}
\left(\frac{M_{0}}{R}\right)_{\rm min}&=&\frac{4\pi}{k^{2}_{f}}\left( -\frac{k^{2}_{f}}{2}P_{\rm eff}R^{2}-\Lambda_f R^{2} \right) \, , \label{minb}  \\
\left(\frac{M_{0}}{R}\right)_{\rm max}&=&\frac{4\pi}{k^{2}_{f}}\left( \frac{4}{3}-\frac{k^{2}_{f}}{6}P_{\rm eff}R^{2}-\Lambda_f R^{2} +\frac{2k^{2}_{f}}{81}\mathcal{D}R^{2}\right) \, .  \notag \\  \label{maxb}
\end{eqnarray}
The nontrivial minimum mass/radius exists when $P_{\rm eff}<-2\Lambda_f/k^{2}_{f}$. This is valid even for $\Lambda_f < 0$.

Mixing conventional massless gravity with massive `gravity' from the covariant interaction term results in the minimum and maximum mass bounds for any static spherical configuration within the framework of the theory. If we take the QCD glueball as the massive spin-2 state to be mixed with the massless graviton, the model will predict the {\it mass gap} of any object composed of particles that couple to the glueball. This does not solve the mass gap problem explicitly, since we assume the glueball mass {\it a priori}. {\it However, the mass gap of the glueball is then transmitted to other particles via the universal interaction of the strong gravity field.} The mass gap is given by the minimum mass and it is proportional to $\mathcal{M}^{2}/k^{2}_{f}$.

That the mass is generated by mixing with strong gravity induced by the glueball and not by chiral symmetry breaking is a remarkable aspect of this mechanism. It is a {\it universal} way to generate mass for a stable static object in the strong interaction. On the other hand, the upper limit represents the maximum mass of the QCD sphere at a given radius~(i.e. maximum density) before it undergoes `gravitational collapse' to form a quark-gluon plasma, which cannot be contained within a static sphere. {\it This strong gravitational collapse is nothing but the deconfinement phase transition in the strong interaction.}  The critical density predicted by this strong gravity model is proportional to $1/k^{2}_{f}$ at the leading order.

However, we may ask, what determines the strong gravity coupling, $k^{2}_{f}$, of the gauge singlet massive `graviton' interaction? As in conventional general relativity, where the Planck mass defines the energy scale at which quantum effects become comparable to those of classical gravity, the analogous mass scale for the strong gravity theory can be used to determine the coupling $k_{f}$ by setting the Compton wavelength of the particle, $\lambda_{\rm C}$, equal to $R$ in Eqn.~(\ref{maxb}) with $P_{\text{eff}},\Lambda_f,\mathcal{D} =0$. We then have
\begin{equation}
k^{2}_{f}=\frac{16\pi}{3}\frac{\hbar}{c^{3}M^{2}_{\rm max}} \, ,
\end{equation}
in the standard units. We define $M_{\rm max}$ to be the corresponding `Planck mass' of the strong gravity. Whereas, for $M > m_{\rm Pl}$, fundamental particles inevitably collapse to form black holes, strongly interacting particles with $M > M_{\rm max}$ inevitably undergo a deconfinement phase transition.  The corresponding length scale, given by a Compton relation, $R_{\rm min} = \hbar/M_{\rm max}c$, may be referred to as the strong gravity `Planck length'.

The mass gap generation mechanism discussed in this paper assumes the glueball mass to be proportional to $\mathcal{M}$ (i.e. the mass mixing term is proportional to $\mathcal{M}^{2}$), while the mass gap itself, given by Eq.~(\ref{minb}), is proportional to $\mathcal{M}^{2}R^{3}/k^{2}_{f}$, yielding
\be
m_{\text{gap}}\simeq\frac{\mathcal{M}^{2}R^{3}}{k^{2}_{f}}=\frac{R^{3}}{R_{f}^{3}}\frac{\mathcal{M}^{2}}{M_{\text{max}}}<\mathcal{M}\frac{R^{3}}{R_{f}^{3}} \, .
\ee
If we set the mass gap equal to the mass of the $\pi$-meson, $m_{\text{gap}} = m_{\pi} \simeq 140$ MeV, $R\simeq R_{f}/2$ and $\mathcal{M}=2$ GeV, the strength of strong gravity becomes
\be
M_{\text{max}} \simeq 3.6 ~\text{GeV}.
\ee
Any quantum particle with a strong gravity interaction will inevitably collapse to form a strong gravity `black hole' once the mass exceeds $M_{\rm{max}}$. The corresponding Hawking temperature of the strong gravity black hole is
\be
T_{\text{max}}=\frac{1}{k^{2}_{f}M_{\text{max}}}=\frac{3M_{\text{max}}}{16\pi} \simeq 0.2 ~\text{GeV} \simeq 2\times10^{12} ~\text{K} \, .
\ee
This is the Hagedorn temperature, i.e. the maximal possible temperature of the nuclear matter, before the phase transition occurs.  Hence, we can identify this temperature with the deconfinement temperature of the hadron. In this picture, strong gravitational collapse {\it is} the deconfinement of the strong nuclear interaction. The strong gravity field $f_{\mu\nu}$ becomes zero/infinity at the `horizon' and reverses sign inside the `black hole'.  We interpret this as the non-existence of the glueball in the deconfined phase.

From the point of view of holographic duality, this looks very familiar. The correspondence between the maximum mass of spherical object in AdS space and the deconfinement temperature of the dual gauge matter is well known~\cite{Witten:1998zw,Danielsson:1999zt,Arsiwalla:2010bt,B5}.
Furthermore, the thermal phase of an AdS black hole is argued to be dual to the thermal phase of deconfined gauge matter living on the boundary of the AdS space. Hence, gravitational collapse in the bulk AdS space corresponds to the deconfinement phase transition of the gauge matter on the AdS boundary. This relationship is holographic in nature since it relates two theories living in different dimensions of spacetime. It is interesting to note similar features of the strong gravity model.

In \cite{B2,B3}, it was argued that the minimum mass bound should be interpreted as the minimum density required for the nuclear matter to maintain its static configuration without evaporating into a hadron gas. However, in light of the strong gravity model presented here, the minimum mass bound in the bulk AdS could also very well be interpreted as corresponding to the minimum mass of {\it any} nuclear particle which is stable under the strong interaction. (Note, however, that such particles may still be unstable with respect to other interactions, such as weak decay, etc.) In this picture, the minimum mass bound in the bulk simply {\it is} the mass gap in the strong interaction on the boundary.

Finally, before concluding this Section, we comment on the requirement that $P_{r}+\rho=0$, which originates from the choice $\Delta = {\rm const.}$ in the `gauge fixing' of the interaction term of the strong gravity Lagrangian. Even though a $U(1)$ charged sphere satisfies this equation of state, generic matter does not obey this condition. In this case, the quantity $\Delta$ becomes physically relevant and we need to allow $\Delta$ to depend on the radial coordinate $r$, setting $\Delta = {\rm const.} \rightarrow \Delta(r)$. We then have
\be
\Delta(r)=e^{\nu+\lambda}=\exp\Big(\int^{r}k^{2}_{f}(P_{r}+\rho)e^{\lambda(r)}r~dr\Big) \, ,
\ee
resulting in an increasing function $\Delta(r)\geq 1$ for a positive $(P_{r}+\rho)$ matter profile. The mass bound analysis above can then be repeated with
\be
\rho_{\rm eff} \equiv \rho+\frac{3\mathcal{M}^{2}}{4\Delta^{3/2}k^{2}_{f}}
\ee
replacing $\rho$ in Eq.~(\ref{masseq}) and by setting $b=0$. The resulting minimum and maximum mass/radius ratio bounds are exactly the same as in Eqns.~(\ref{minb}) and (\ref{maxb}) under the replacement $\Lambda_f \to \Lambda_f^{\rm (min)}, \Lambda_f^{\rm (max)}$ where
\begin{eqnarray*}
\Lambda_f^{\rm (min)} &=& \frac{3\mathcal{M}^{2}}{4\Delta^{3/2}(r=R)} \, , \notag \\
\Lambda_f^{\rm (max)} &=& \frac{3\mathcal{M}^{2}}{4\Delta^{3/2}(r=0)}=\frac{3\mathcal{M}^{2}}{4} \, , \notag\\
\end{eqnarray*}
respectively. The existence of mass gap is generic in this kind of model.

For convenience, here and henceforth we rewrite the metric in a rescaled coordinate $r'= r/\sqrt{3/2}$ and redefine $\Lambda_{f}\to \Lambda_{f}/3$ so that our metric is in the conventional form.

\section{Quantum mechanical implications of the classical mass/radius ratio bounds in strong gravity} \label{sect5}

In the present Section, we investigate the quantum mechanical implications of the mass/radius bounds in the strong gravity model. We begin with a brief discussion of the quantum implications of mass bounds in conventional general relativity, in the presence of a cosmological constant $\Lambda \neq 0$, before extending these to the strong gravity case via the substitutions $G \rightarrow G_f$, $\Lambda \rightarrow \Lambda_f$.

\subsection{Quantum mass bounds in standard general relativity} \label{sect5.1}

For fundamental particles, viewed as stable compact objects, the radius of an uncharged particle may be identified with the Compton wavelength $\lambda_{\rm C}$, or reduced Compton wavelength, $k_{\rm C}^{-1}$, given by
\be \label{Compton}
\lambda_{\rm C} = \frac{h}{Mc} \iff k_{\rm C}^{-1} = \left(\frac{2\pi}{\lambda_{\rm C}}\right)^{-1} = \frac{\hbar}{Mc} \, .
\ee
For order of magnitude relations, we use these two expressions interchangeably from here on. The combination of Eqs. (\ref{DE_dens}) and (\ref{Compton}) then implies the existence of minimum mass for a stable, charge-neutral, quantum mechanical and gravitating compact object,
\be \label{m_Lambda}
M \gtrsim M_{\Lambda} = \sqrt{m_{\rm Pl}m_{\rm dS}} = 6.833 \times 10^{-36} \rm g \, ,
\ee
where, for future reference, we define the (reduced) Planck mass and length scales
$m_{\rm Pl} = \sqrt{\hbar c/G} =  2.176 \times 10^{-5} \ {\rm g}$, and
$l_{\rm Pl} = \sqrt{\hbar G/c^3} = 1.616 \times 10^{-33} \ {\rm cm}$, respectively,
and the (reduced) first/second de Sitter mass and length scales, denoted using unprimed/primed quantities, respectively, as
\begin{eqnarray} \label{dS_scales}
m_{\rm dS} &=& \frac{\hbar}{c}\sqrt{\frac{\Lambda}{3}} =2.145 \times 10^{-66} \ {\rm g} ,\nonumber\\
m_{\rm dS}' &=& \frac{c^2}{G}\sqrt{\frac{3}{\Lambda}} =2.210 \times 10^{56} \ {\rm g} \, ,
\nonumber\\
l_{\rm dS} &=& \sqrt{\frac{3}{\Lambda}} = 1.641 \times 10^{28} \ {\rm cm} \, ,
\nonumber\\
l_{\rm dS}' &=& \frac{\hbar c}{G}\sqrt{\frac{\Lambda}{3}} = 1.593 \times 10^{-94} \ {\rm cm} \, .
\end{eqnarray}

Note that, in previous work \cite{B1,B2,B3}, the de Sitter scales defined in Eq.~(\ref{dS_scales}) were also referred to as the first and second Wesson scales, following the pioneering work \cite{Wesson:2003qn}.
The physical interpretations of these scales are discussed in detail in \cite{B1}. For now, we simply note that the first and second de Sitter scales are related via
\be \label{dS_T-duality}
m_{\rm dS}' = \frac{m_{\rm Pl}^2}{m_{\rm dS}} \, \quad l_{\rm dS}' = \frac{l_{\rm Pl}^2}{l_{\rm dS}} \, ,
\ee
and that the numerical value of the minimum mass, $m_{\Lambda} \simeq 10^{-3} \rm eV$ is consistent with current experimental bounds on the mass of the electron neutrino, the lightest known neutral particle, obtained from the Planck satellite data, $m_{\nu} \leq 0.23 \ \rm eV$ \cite{Planckresults}. To within numerical factors of order unity, $m_{\Lambda}$ is also the {\it unique} mass scale for which the Compton radius of the particle is equal to its gravitational turn-around radius,
\begin{eqnarray}  \label{turn-round_radius}
r_{\rm grav} = 2^{-1/3}(l_{\rm dS}^2r_{\rm S})^{1/3} \, , \quad (r_{\rm S} = 2GM/c^2) \, ,
\end{eqnarray}
in the presence of dark energy \cite{Bhattacharya:2016vur,LakePaterek1,Lake1}. This represents the radius beyond which the repulsive effects of the background dark energy dominate over the attractive force of canonical gravity. Eq.~(\ref{turn-round_radius}) may be obtained by considering the Newtonian limit of general relativity for the Schwarzschild-de Sitter metric, which gives rise to an effective Newtonian potential of the form
\begin{eqnarray}  \label{Lambda_Modified_Newtonian_potential}
\Phi(r) = -\frac{GM}{r} - \frac{\Lambda c^2}{6}r^2 \, .
\end{eqnarray}
This, in turn, gives the effective gravitational field strength \cite{Hobson:2006se}
\begin{eqnarray}  \label{Lambda_Modified_Newtonian_g}
\vec{g}(r) = -\vec{\nabla}\Phi(r) = \left(-\frac{GM}{r^2} + \frac{\Lambda c^2}{3}r\right) \hat{\vec{r}} \, ,
\end{eqnarray}
which changes sign at $r=r_{\rm grav}$. However, to within numerical factors of order unity, the expression (\ref{turn-round_radius}) remains rigorously valid in full general relativity \cite{Bhattacharya:2016vur}. This gives a neat way of reinterpreting the stability bound (\ref{min_dens_Lambda>0}). If the quantum mechanical (i.e. Compton) radius of the particle lay outside its gravitational turn-around radius, it would clearly be unstable due to dark energy repulsion.
Interestingly, we may also ask, for what mass is the turn-around radius equal to the Schwarzschild radius? The answer is $M \simeq m_{\rm dS}' = (c^2/G)\sqrt{3/\Lambda}$, which is comparable to the present day mass of the Universe \cite{B1,B2}.

\subsection{Quantum mass bounds in strong gravity}  \label{sect5.2}

\subsubsection{Mass bounds for neutral particles}  \label{sect5.2.1}

The Planck mass and length scales are obtained by equating the Compton wavelength of a quantum mechanical particle with the Schwarzschild radius induced by its classical gravitational field (ignoring numerical factors of order unity). In the strong gravity model, an analogous construction using the Schwarzschild radius of the $f_{\mu\nu}$ field  gives
\begin{eqnarray} \label{SG_Planck_scales}
m_{\rm sP} &=& \sqrt{\frac{\hbar c}{G_f}} = 2.176 \times 10^{-24} \ {\rm g} \, ,
\nonumber\\
l_{\rm sP} &=& \sqrt{\frac{\hbar G_f}{c^3}} = 1.616 \times 10^{-14} \ {\rm cm} \, ,
\end{eqnarray}
where $m_{\rm sP}$ and $l_{\rm sP}$ denote the strong gravity Planck mass and strong gravity Planck length, respectively. Note that $m_{\rm sP}$ and $l_{\rm sP}$ are equivalent to $M_{\rm max}$ and $R_{\rm min}$, defined in Sec. \ref{sect4}. We here relabel these quantities for the sake of easy comparison with results from standard general relativity. Likewise, analogues of the de Sitter scales may be obtained by replacing $G \rightarrow G_f$ and $\Lambda \rightarrow \Lambda_f$ in Eq. (\ref{dS_scales}). In addition, based on purely dimensional arguments, we may define two additional mass scales, and their corresponding lengths, by mixing and matching $\left\{G,\Lambda,G_f,\Lambda_f\right\}$.

To investigate the physical meaning (if any) of these scales, and of other mass/length scales constructed using the strong gravity model parameters, we must first consider the physical interpretation of the mass term $\mathcal{M}^2$ in the mixing Lagrangian $\mathcal{L}_{fg}$. This appears in the strong gravity field equations in combination with the `geometric' parameter $\Delta$, through the definition of the strong cosmological constant, $\Lambda_f =\mathcal{M}^2/4\Delta ^{3/2}$ (\ref{strong_Lambda}). By analogy with Eq. (\ref{DE_dens}), we define the energy density associated with $\Lambda_f$ as
\be \label{strong_DE_dens}
\rho_{\Lambda_f} = -p_{\Lambda_f}/c = \frac{\Lambda_f c^2}{8\pi G_f} \, .
\ee
This may be related to the equation of state for deconfined quark matter, obtained from perturbation theory in QCD, as follows. Neglecting quark masses in the first order perturbation, the relation between the pressure and energy density of nuclear matter is given by
\be \label{bag_model_EOS}
p/c =(\rho - 4B)/3 \, ,
\ee
where $B \simeq 2 \times 10^{14} \rm gcm^{-3}$ is the difference in energy density between the perturbative and the non-perturbative QCD vacuums, and is of the order of the nuclear density. This model is known as the MIT `bag' model and the constant $B$ is called the bag constant. When nuclear matter is compressed to sufficiently high density, a phase transition is thought to occur which converts confined hadronic matter into free, three-flavor (strange) quark matter. The collapse of the quark fluid is described by the bag model equation of state (\ref{bag_model_EOS}).

In \cite{sg3}, it was already pointed out in that the QCD bag constant effectively resembles a cosmological constant for strongly interacting matter. Qualitatively at least, it is not difficult to understand how the effective potential of the strong gravity model mimics the effective (`bag-type') potential obtained from QCD. The Newtonian limit of the Schwarzschild-de Sitter type metric for the $f_{\mu\nu}$ field gives
\begin{eqnarray}  \label{Lambda_Modified_Newtonian_potential_s}
\Phi_{\rm s}(r) = -\frac{G_f M}{r} - \frac{\Lambda_f c^2}{6}r^2 \, ,
\end{eqnarray}
\begin{eqnarray}  \label{Lambda_Modified_Newtonian_g_s}
\vec{g}_{\rm s}(r) = -\vec{\nabla}\Phi_s(r) = \left(-\frac{G_f M}{r^2} + \frac{\Lambda_f c^2}{3}r\right) \hat{\vec{r}} \, ,
\end{eqnarray}
by analogy with Eqs. (\ref{Lambda_Modified_Newtonian_potential})-(\ref{Lambda_Modified_Newtonian_g}). However, here, the $f_{\mu\nu}$ `vacuum' corresponds to the presence of strongly interacting matter and the `cosmological constant' $\Lambda_f$ is generated by the interaction term in the strong gravity Lagrangian, $\mathcal{L}_{fg}$. Hence, the strong force is attractive on small scales, $r \leq r_{\rm grav(s)}$, where $r_{\rm grav(s)}$ is the strong gravity turn-around radius,
\begin{eqnarray}  \label{SGturn-round_radius}
r_{\rm grav(s)} = 2^{-1/3}(l_{\rm sd}^2r_{\rm sS})^{1/3} \, , \quad (r_{\rm sS} \equiv 2G_fM/c^2) \, ,
\end{eqnarray}
repulsive on intermediate scales, $r_{\rm grav(s)} \leq r \leq R$, and quickly tends to zero in the true vacuum ($r \gtrsim R$), where the density of the strongly interacting matter also falls quickly to zero. Here, $r_{\rm sS}$ denotes the string gravity Schwarzschild radius. 

However, {\it if} we may identify the outer radius of the strongly interacting `particle' with the strong gravity turn-around radius $R \simeq r_{\rm grav(s)}$, the repulsive phase is never realised. The effective potential is strongly attractive over short distances and is (in principle) capable of countering the effects of electrostatic repulsion if the matter is also charged.  It then vanishes at $R \simeq r_{\rm grav(s)}$ and remains zero outside the particle. Furthermore, if the analysis presented above is modified to include $r$-dependence in the geometric parameter $\Delta$ in Eqn.~(\ref{lambeq}), i.e. such that
\be
\Delta \rightarrow \Delta(r)\propto r^{2/3} \, , 
\ee
we may generate a {\it confining potential}, $\Phi_{\rm s} \propto r$, for $r \gtrsim r_{\rm grav(s)} \simeq R$. This possibility is discussed further in Sec. \ref{sect6}. In the present section, we assume an effective potential of the form (\ref{Lambda_Modified_Newtonian_g_s}), which holds up to $r \simeq r_{\rm grav}$, given by (\ref{SGturn-round_radius}). This allows us to treat the `bag' of the MIT bag model within the context of the strong gravity theory.

Technically, since $B$ has dimensions of density, it plays the role of the `vacuum' energy density associated with $\Lambda_f$ (\ref{strong_DE_dens}), which is nonzero {\it inside} the strongly interacting particles. We note also that, for $\rho = B \simeq {\rm const.}$, we have $\rho = -p/c \simeq {\rm const.}$, a dark energy-type equation of state. Hence, we may identify
\bea \label{B_vs_Lambda_f}
B \simeq 2 \times 10^{14} \rm gcm^{-3} &\equiv& \rho_{\Lambda_f} = \frac{\Lambda_f c^2}{8\pi G_f} \, ,
\nonumber\\
&=& \rho_{\Lambda} \times \frac{\Lambda_f }{\Lambda}\frac{G}{G_f} \, ,
\eea
where $\rho_{\Lambda} = \Lambda c^2/(8\pi G_f) = 5.971 \times 10^{-30} {\rm gcm^{-3}}$, $\Lambda = 1.114 \times 10^{-52} \ {\rm cm^{-3}}$ are the current best fit values obtained from cosmological observations \cite{Planckresults,Betoule:2014frx}. For $G_f \simeq 10^{38}G$ (\ref{Dirac'sNo.}), this implies
\be  \label{Lambda_f_estimate}
\Lambda_f \simeq 3.733 \times 10^{25} \ {\rm cm^{-2}} \simeq 3.351 \times 10^{81}\Lambda \, .
\ee
For this value of $\Lambda_f$ we obtain, by analogy with Eq. (\ref{dS_scales}),
\begin{eqnarray} \label{SG_dS_scales}
m_{\rm sd} &=& \frac{\hbar}{c}\sqrt{\frac{\Lambda_f}{3}} = 1.241 \times 10^{-25} \ {\rm g} \, ,
\nonumber\\
m_{\rm sd}' &=& \frac{c^2}{G_f}\sqrt{\frac{3}{\Lambda_f}} = 3.818 \times 10^{-23} \ {\rm g} \, ,
\nonumber\\
l_{\rm sd} &=& \sqrt{\frac{3}{\Lambda_f}} = 2.835 \times 10^{-13} \ {\rm cm} \, ,
\nonumber\\
l_{\rm sd}' &=& \frac{\hbar c}{G_f}\sqrt{\frac{\Lambda_f}{3}} = 9.214 \times 10^{-16} {\rm cm} \, ,
\end{eqnarray}
where the subscript `sd' stands for `strong-de Sitter'. Remarkably, we note that, for $G_f \simeq 10^{38}G$ (\ref{Dirac'sNo.}), the strong gravity Planck and de Sitter scales are {\it approximately} equal, 
\begin{eqnarray}
m_{\rm sP} \simeq m_{\rm sd} \simeq 10^{-24}-10^{-25} \ {\rm g} \ \ \ (G_f \simeq 10^{38}G) \, .
\end{eqnarray}
As we shall see, this has important implications for the model as an effective theory, able to mimic confinement. 

Additional mixing and matching of $\left\{G,\Lambda,G_f,\Lambda_f\right\}$ also yields
\begin{eqnarray} \label{Salam_scales}
m_{\rm Sa} &=& \frac{m_{\rm Pl}^2}{m_{\rm sd}} = \frac{c^2}{G}\sqrt{\frac{3}{\Lambda_f}} = 3.817 \times 10^{15} \ {\rm g} \, ,
\nonumber\\
m_{\rm Sa}' &=& \frac{m_{\rm sP}^2}{m_{\rm dS}} = \frac{c^2}{G_f}\sqrt{\frac{3}{\Lambda}} = 2.210 \times 10^{18} \ {\rm g} \, ,
\nonumber\\
l_{\rm Sa} &=& \frac{l_{\rm Pl}^2}{l_{\rm sd}} = \frac{\hbar c}{G}\sqrt{\frac{\Lambda_f}{3}} = 9.214 \times 10^{-54} \ {\rm cm} \, ,
\nonumber\\
l_{\rm Sa}' &=& \frac{l_{\rm sP}^2}{l_{\rm dS}} = \frac{\hbar c}{G_f}\sqrt{\frac{\Lambda}{3}} = 1.592 \times 10^{-59} \ {\rm cm} \, ,
\end{eqnarray}
based on purely dimensional arguments. We christen these the first and second Salam mass/length scales, though their physical interpretations are not investigated here as such an analysis lies beyond the scope of the present work. 

Instead, in the analysis that follows, we focus on the strong gravity analogues of mass/length scales which are well defined and understood in standard general relativity, and on additional mass/length scales that may be derived from them using quantum gravity arguments. These include the strong gravity Planck scales (\ref{SG_Planck_scales}), strong gravity de Sitter scales (\ref{SG_dS_scales}), the strong gravity Schwarzschild and turn-around radii (\ref{SGturn-round_radius}) and the analogues of the charge-neutral an charged particle mass bounds, $M \gtrsim M_{\Lambda}$ (\ref{m_Lambda}) and $M \gtrsim M_{Q}$ (\ref{m_Q}), obtained by replacing $G \rightarrow G_f$ and $\Lambda \rightarrow \Lambda_f$. In addition, we consider a modified charge-neutral bound, $M \gtrsim \tilde{M}_{\Lambda}$, which corresponds to replacing $G \rightarrow G_f$, $\Lambda \rightarrow \Lambda_f$ and identifying the compact radius $R$ with the scattering radius of the particle, rather than its Compton wavelength. As we shall see, this allows us to recover the mass of the neutron as the mass of the lightest, stable, compact, charge-neutral and strongly interacting quantum mechanical particle in nature, according the strong gravity theory.


In strong gravity, the analogue of the stability condition for neutral particles in general relativity with $\Lambda > 0$, Eq. (\ref{min_dens_Lambda>0}), is
\be \label{SG_min_dens_Lambda_f>0}
\frac{2G_fM_0}{c^2R}\geq \frac{\Lambda_f}{6}R^2 \iff \rho \geq \rho _{\rm min(s)} := \frac{3M_0}{4\pi R^3}\geq \frac{\Lambda_f c^2}{16\pi G_f} \, ,
\ee
where $M_0$ is the bare mass of the hadron. (However, from here on, we simply use $M_0 \rightarrow M$ to denote the bare mass.) This follows directly from Eq. (\ref{76}) for the quasi-anisotropic case, $\mathcal{D}=0$.

Even though the bounds obtained in previous sections are derived from the {\it classical} strong gravity field equations, we would like to extend them to the quantum mechanical regime. Thus, we consider the situation where the minimum bound is saturated by a `classical' particle that is equally quantum mechanical, in the sense that its classical size is equal to its Compton quantum wavelength. The particle will become `purely' quantum if its classical radius becomes even smaller. Setting $R = k_{\rm C}^{-1}$, where $k_{\rm C}^{-1}$ denotes the reduced Compton wavelength (\ref{Compton}), Eqn.~(\ref{SG_min_dens_Lambda_f>0}) then gives
\be \label{m>m_Lambda_f}
M \gtrsim M_{\Lambda_f} = \sqrt{m_{\rm sP}m_{\rm sd}} = 5.197 \times 10^{-25} \ {\rm g} \, .
\ee
This may also be obtained, to within numerical factors of order unity, by equating the Compton scale with the strong gravity turn-around radius, defined by Eq. (\ref{SGturn-round_radius}).

Hence, according to the strong gravity theory, combined with elementary quantum mechanics, $M_{\Lambda_f} \simeq 10^{-25}$g should correspond to the mass of the {\it lightest possible stable, compact, charge neutral, strongly interacting and quantum mechanical particle found in nature.} With this in mind, we note that this is {\it almost} equal to the neutron rest mass $m_n \simeq 1.675 \times 10^{-24}$ g, though a discrepancy of around one order of magnitude remains. That said, also we note that the neutron is {\it not} a fundamental particle. Therefore it is unclear whether $R$ in Eq. (\ref{SG_min_dens_Lambda_f>0}) should be identified with $\lambda_n = \lambda_{\rm C}(m_n) = 2.100 \times 10^{-14}$ cm -- here we estimate the {\it reduced} Compton wavelength but use the standard (ambiguous) notation -- or with some other measure of the neutron radius.

In particular, we may instead consider the neutron radius obtained from scattering cross section data, $\sigma_n = \pi r_n^2 \simeq 10^{-24} \ {\rm cm^{-2}}$, yielding $r_n \simeq 5.642 \times 10^{-13} \ {\rm cm} \simeq 26.864 \times \lambda_n$. Identifying the radius $R$ in Eq. (\ref{SG_min_dens_Lambda_f>0}) with the scattering radius of the particle $r_{\rm scat}$,
\be \label{scattering_CS}
R = r_{\rm scat} \equiv \chi_{\rm scat} \lambda_{\rm C} \, , \quad (\chi_{\rm scat} < 1) \, ,
\ee
yields an alternative estimate of the lightest neutral hadron in the strong gravity theory, which we denote $\tilde{M}_{\Lambda_f}$. This is given by
\be \label{m>m_Lambda_f*}
M \gtrsim \tilde{M}_{\Lambda_f} = \chi_{\rm scat}^{3/4}\sqrt{m_{\rm sP}m_{\rm sd}} = 5.197 \times 10^{-25}  \chi_{\rm scat}^{3/4} \ {\rm g} \, .
\ee
For $\chi_{\rm scat} \simeq 26.864$, the relevant value for the neutron, this yields
\bea \label{m>m_Lambda_f**}
\tilde{M}_{\Lambda_f} &=& 11.800 \times m_{\Lambda_f} = 6.132 \times 10^{-24} \ {\rm g}
\nonumber\\
&\simeq& m_n \simeq 1.675 \times 10^{-24} \ {\rm g} \, .
\eea

Finally, we note that, due to the {\it approximate} numerical coincidence of $m_{\rm sP}$ and $m_{\rm sd}$, we have that
\bea \label{equivs}
\lambda_{\rm C}(M_{\Lambda_f}) &\simeq& r_{\rm sS}(M_{\Lambda_f}) \simeq r_{\rm grav(s)}(M_{\Lambda_f})
\nonumber\\
&\simeq& l_{\rm sP} \simeq l_{\rm sd}
\nonumber\\
&\sim& \mathcal{O}(10^{-13})-\mathcal{O}(10^{-14}) \ {\rm cm} \, ,
\eea
and similar relations hold for $M = \tilde{M}_{\Lambda_f}$. This justifies our earlier assumption that the effective potential (\ref{Lambda_Modified_Newtonian_g_s}) holds up to $R \simeq r_{\rm grav(s)}$ (\ref{SGturn-round_radius}).

For convenience, we denote the {\it reduced} Compton scales associated with $M_{\Lambda}$, $M_{\Lambda_f}$ and $\tilde{M}_{\Lambda_f}$ as
\bea
l_{\Lambda} = \sqrt{l_{\rm Pl}l_{\rm dS}} = 5.150 \times 10^{-3} \ {\rm cm} \,
\nonumber\\
\eea
\be
l_{\Lambda_f} = \sqrt{l_{\rm sP}l_{\rm sd}} \simeq 6.769 \times 10^{-14} \ {\rm cm} \, ,
\ee
and
\be
\tilde{l}_{\Lambda_f} = \chi_{\rm scat}^{-3/4}\sqrt{l_{\rm sP}l_{\rm sd}} \simeq 5.737 \times 10^{-15} \ {\rm cm} \, ,
\ee
respectively, from now on.

\subsubsection{Mass bounds for charged particles}  \label{sect5.2.2}

Having considered neutral particles, we now try to combine classical stability bounds for {\it charged} strongly interacting fluid spheres with quantum mechanics. To this end, we now (briefly) review the derivations of Eqs. (\ref{m_Q})-(\ref{Lambda_vs_re}) presented in references \cite{B2,LakePaterek1,Lake1}. (Note that Eq. (\ref{Lambda_vs_re}) was also derived, using different methods, in \cite{Nottale1993,Beck:2008rd} and that its cosmological implications were investigated in \cite{Wei:2016moy}, while similar expressions, equivalent to replacing $m_e/\alpha_e \leftrightarrow m$, were obtained in \cite{Zel'dovich:1968zz} and \cite{Funkhouser:2005hp}.)

As we shall see, the approach taken in \cite{B2,LakePaterek1,Lake1}, based on hypothetical minimum length uncertainty relations (MLURs), may be readily extended to the strong gravity theory, leading to expressions analogous to Eqs. (\ref{m_Q})-(\ref{Lambda_vs_re}), but with $G \rightarrow G_f$ and $\Lambda \rightarrow \Lambda_f$. Having obtained these, we again identify the energy density associated with $\Lambda_f$, $\rho_{\Lambda_f} \equiv \Lambda_fc^2/(8\pi G_f)$, with the `bag constant' of the MIT bag model, $B \simeq 2 \times 10^{14} \ {\rm gcm^{-3}}$. This, in turn, allows us to obtain a numerical estimate of the minimum mass of a {\it stable, compact, charged, strongly interacting and quantum mechanical particle}. For $Q = \pm 2e/3$, this is found to be of the same order of magnitude as the masses of the lightest known particle of this form, i.e. the mass of the up quark, which is believed to lie in the range 1.7-3.3 MeV.

We emphasise that, when estimating the the mass of the lightest stable, charge neutral and strongly interacting quantum mechanical particle, we {\it expected} to obtain an estimate of the neutron mass $m_n$, whose density is $\rho_n \simeq B$. By contrast, when considering the lightest possible {\it charged} and strongly interacting quantum mechanical particle, we expect to obtain an estimate of the lightest known quark mass. This is because there are no known {\it fundamental}, charge-neutral, and strongly interacting particles in nature, whereas fundamental charged and strongly interacting particles (i.e. quarks) do exist. However, in considering the mass of a {\it free} quark, we must consider the point at which it becomes unconfined, and identify this with the `strong dark energy' density after the phase transition to the quark-gluon plasma.

In \cite{Ng:1993jb,Ng:1994zk}, an MLUR of the form
\bea \label{MLUR}
\Delta x_{\rm total}(\Delta v,r,M) &\gtrsim& \frac{\lambda_{\rm C}}{2}\frac{c}{\Delta v} + \alpha'\frac{\Delta v}{c}r + \beta \frac{l_{\rm Pl}^2}{\lambda_{\rm C}}
\nonumber\\
&\gtrsim& \sqrt{2\alpha'}\sqrt{\lambda_{\rm C}r} + \beta\frac{l_{\rm Pl}^2}{\lambda_{\rm C}} \, ,
\eea
where $\alpha'$, $\beta = {\rm const.}$, was proposed. Here, $\Delta x_{\rm total}$ represents the minimum possible uncertainty in the position of a `probe' particle, which is used to measure a distance $r = ct$ by means of the emission and reabsorption of a photon. Thus, it is equal to the minimum possible uncertainty in the measurement of the probe distance $r$. The first term on the top line of Eq. (\ref{DE-UP}) is the standard Heisenberg term, rewritten using the relations $\Delta p = M\Delta v$ and $\lambda_{\rm C} \simeq k_{\rm C}^{-1} = \hbar/(Mc)$, whereas the second represents a recoil term, due to the emission of the photon \cite{Salecker:1957be}. The third is the `gravitational uncertainty', which is assumed to be of the order of the Schwarzschild radius $r_{\rm S} \simeq l_{\rm Pl}^2/\lambda_{\rm C}$ \cite{Ng:1993jb,Ng:1994zk}. The second line is obtained by minimising the first with respect to $\Delta v$, giving
\be \label{Delta_v_max}
\Delta v \lesssim \Delta v_{\rm max} \simeq \frac{1}{\sqrt{2\alpha'}}\sqrt{\frac{\lambda_{\rm C}}{r}}c \, .
\ee
Minimising the expression on the second line of Eq. (\ref{MLUR}) with respect to $M$ then yields
\begin{eqnarray} \label{MLUR-2}
&&M = \left(\frac{\alpha'}{2\beta^2}\right)^{1/3}\left(\frac{r}{l_{\rm Pl}}\right)^{1/3}m_{\rm Pl}
\nonumber\\
&\iff& r \equiv r_{\rm min} = \left(\frac{2\beta^2}{\alpha'}\right)\left(\frac{M}{m_{\rm Pl}}\right)^3 l_{\rm Pl}\, ,
\end{eqnarray}
and hence
\begin{eqnarray} \label{MLUR-3}
(\Delta x_{\rm total})_{\rm min} \simeq 3\left(\frac{\alpha'\beta}{2}\right)^{1/3}(l_{\rm Pl}^2r)^{1/3} \, .
\end{eqnarray}
The $M$ in Eq. (\ref{MLUR-2}) represents the optimum mass for the probe particle. This yields the minimum possible uncertainty in the measurement of the probe distance $r$, given by Eq. (\ref{MLUR-3}). The probe distance which may be measured with minimum uncertainty, denoted $r_{\rm min}$, is defined via Eq. (\ref{MLUR-2}).

The canonical quantum part of the MLUR (\ref{MLUR}), $(\Delta x_{\rm canon.})_{\rm min} \gtrsim \sqrt{\lambda_{\rm C}r}$, was originally derived by Salecker and Wigner using the gedanken experiment considered above (neglecting the particle's self-gravity) \cite{Salecker:1957be} but may also be derived more rigorously by directly solving the Sch{\" o}dinger equation in the Heisenberg picture, before setting $t = r/c$ \cite{Calmet:2004mp,Calmet:2005mh}. Though derived via different means, an MLUR of the form (\ref{MLUR-3}) was originally obtained by K{\' a}rolyh{\' a}zy, under the assumption of {\it asymptotically flat space}, in \cite{Karolyhazy:1966zz,KFL}. In most of the existing quantum gravity literature, the constants $\alpha'$ and $\beta$ are assumed to be of order unity, $\alpha'$, $\beta \sim \mathcal{O}(1)$ \cite{Hossenfelder:2012jw,Garay:1994en}.

However, in \cite{B2,LakePaterek1,Lake1}, it is argued that the introduction of a constant dark energy density, i.e. $\Lambda >0$, and, hence, the existence of a de Sitter horizon, $l_{\rm dS} = \sqrt{3/\Lambda}$, implies a fundamental modification of the MLUR (\ref{MLUR}), equivalent to the substitution
\be \label{beta'_subs}
\beta = {\rm const.} \rightarrow \beta(r) = \beta' \frac{l_{\rm dS}}{r} \, ,
\ee
where $\beta' = {\rm const.} \sim \mathcal{O}(1)$. Equation (\ref{MLUR}) then becomes
\bea \label{DE-UP}
\Delta x_{\rm total}(\Delta v,r,M) &\gtrsim& \frac{\lambda_{\rm C}}{2}\frac{c}{\Delta v} + \alpha'\frac{\Delta v}{c}r + \beta' \frac{l_{\rm Pl}^2l_{\rm dS}}{\lambda_{\rm C}r}
\nonumber\\
&\gtrsim& \sqrt{2\alpha'}\sqrt{\lambda_{\rm C}r} + \beta'\frac{l_{\rm Pl}^2l_{\rm dS}}{\lambda_{\rm C}r} \, ,
\eea
and the analogues of Eqs. (\ref{MLUR-2})-(\ref{MLUR-3}) are
\begin{eqnarray} \label{DE-UP-2}
&& M = \left(\frac{\alpha'}{2\beta'^2}\right)^{1/3} \frac{r}{(l_{\rm Pl}^2l_{\rm dS})^{1/3}} (m_{\rm Pl}^2m_{\rm dS})^{1/3}
\nonumber\\
&\iff& r \equiv r_{\rm min} \equiv \left(\frac{2\beta'^2}{\alpha'}\right)^{1/3}(l_{\rm Pl}l_{\rm dS}^2)^{1/3}\frac{M}{m_{\rm Pl}} \, ,
\end{eqnarray}
and
\begin{eqnarray} \label{DE-UP-3}
(\Delta x_{\rm total})_{\rm min} \simeq R_{\rm cell} \equiv 3\left(\frac{\alpha'\beta'}{2}\right)^{1/3}(l_{\rm Pl}^2l_{\rm dS})^{1/3} \, ,
\end{eqnarray}
respectively. Here, $R_{\rm cell}$ represents the linear dimension associated with a fundamental `cell' within the de Sitter horizon, yielding a holographic relation between the number of degrees of freedom in the bulk and on the boundary \cite{B2}.

In \cite{LakePaterek1} it is also shown that the dark energy-modified MLUR, dubbed the `dark energy uncertainty principle' or DE-UP for short, is consistent with the minimum-mass bound $M \gtrsim M_{\Lambda}$, obtained independently in \cite{B1}. Requiring every (potentially) observable length scale in the DE-UP, i.e. $r$, $(\Delta x_{\rm canon.})_{\rm min} \gtrsim \sqrt{\lambda_{\rm C}r}$ and $\Delta x_{\rm grav} \simeq \beta'l_{\rm Pl}^2l_{\rm dS}/(\lambda_{\rm C}r)$, to be super-Planckian leads naturally to Eq. (\ref{m_Lambda}). In addition, since Eq. (\ref{DE-UP}) is invariant under simultaneous rescalings of the form
\begin{eqnarray} \label{rescalings}
\Delta v &\rightarrow& \alpha_{\rm Q}^{-1}\Delta v \, ,
\nonumber\\
M &\rightarrow& \alpha_{\rm Q}M \, ,
\nonumber\\
r &\rightarrow& \alpha_{\rm Q}r \, ,
\end{eqnarray}
where $\alpha_{\rm Q} > 0$ is a positive real parameter, the minimum uncertainty (\ref{MLUR-3}) may {\it also} be obtained for rescaled values of $\Delta v_{\rm max}$ and $M$, obtained by applying (\ref{rescalings}) to Eqs. (\ref{Delta_v_max}) and (\ref{DE-UP-2}).

These results may be combined with Bekenstein's relation for the for the stability of charged, self-gravitating fluid spheres (Eq. (\ref{Buch_Q_Lambda}) in the limit $R \lesssim r_{\rm grav}(M)$), by identifying $(\Delta x_{\rm total})_{\rm min} \simeq R \gtrsim Q^2/(Mc^2)$. This is equivalent to assuming that the particle {\it simultaneously saturates both the classical and quantum stability bounds}, and allows us to solve the resulting equations explicitly, yielding
\begin{eqnarray}  \label{M_crit}
M \simeq \alpha_{\rm Q}M_{\rm T} \simeq \alpha_{\rm Q}(m_{\rm Pl}^2m_{\rm ds})^{1/3}  \, ,
\end{eqnarray}
together with
\begin{eqnarray}  \label{oom-relns-2}
(\Delta x_{\rm total})_{\rm min} \simeq R \simeq \alpha_{\rm Q} \lambda_{\rm C} \simeq r_{\rm min}/\alpha_{\rm Q} \, ,
\end{eqnarray}
where $\alpha_{\rm Q} = Q^2/q_{\rm Pl}^2$ (\ref{alpha_Q}). Evaluating Eq. (\ref{M_crit}) for $Q^2 = e^2$ and reinserting the inequality arising from Eq. (\ref{Buch_Q_Lambda}) then yields Eqs. (\ref{m>me_bound})-(\ref{Lambda_vs_re}), given in the Introduction.

For strongly interacting particles, the relevant horizon is the strong de Sitter radius $l_{\rm sd}$, defined in Eq. (\ref{SG_Planck_scales}), and the relevant `Planck scale' is $l_{\rm sP}$, given in Eq. (\ref{SG_dS_scales}). Modifying Eq. (\ref{DE-UP}) to incorporate the `gravitational uncertainty' of the {\it strong gravity metric} $f_{\mu\nu}$, by making the substitutions
$l_{\rm Pl} \rightarrow l_{\rm sP}$, $l_{\rm dS} \rightarrow l_{\rm sd}$, or equivalently $G \rightarrow G_f$, $\Lambda \rightarrow \Lambda_f$, yields
\bea \label{SG_DE-UP}
\Delta x_{\rm total(s)}(\Delta v,r,M) &\gtrsim& \frac{\lambda_{\rm C}}{2}\frac{c}{\Delta v} + \alpha'\frac{\Delta v}{c}r + \beta' \frac{l_{\rm sP}^2l_{\rm sd}}{\lambda_{\rm C}r}
\nonumber\\
&\gtrsim& \sqrt{\lambda_{\rm C}r} + \beta'\frac{l_{\rm sP}^2l_{\rm sd}}{\lambda_{\rm C}r} \, .
\eea
Combining this with the analogue of the generalised Buchdahl-Bekenstein bound for the strongly interacting fluid sphere,
\bea \label{SG_Buch_Q_Lambda}
M &\geq& \frac{3}{4}\frac{Q^2}{Rc^2} + \frac{c^2}{G_f}\frac{\Lambda_f R^3}{6}
\nonumber\\
&\gtrsim& \frac{3}{4}\frac{Q^2}{Mc^2} \, , \quad \left(R \lesssim r_{\rm grav}(M) \simeq (l_{\rm sd}^2r_{\rm sS})^{1/3}\right) \, ,
\eea
gives the radius of a holographic strong gravity metric `cell' as
\begin{eqnarray} \label{SG_R_cell}
R_{\rm cell(s)} \simeq (l_{\rm sP}^2l_{\rm sd})^{1/3}  \simeq 4.120 \times 10^{-14} \ {\rm cm} \, .
\end{eqnarray}
We interpret the associated mass scale,
\begin{eqnarray} \label{SG_m_Q}
M \gtrsim \alpha_{\rm Q}M_{\rm sT} &\equiv& \alpha_{\rm Q}(m_{\rm sP}^2m_{\rm sd})^{1/3}
\nonumber\\
&\simeq& (Q^2/\hbar c)(\hbar^2\sqrt{\Lambda_f}/G_f)^{1/3} \, ,
\end{eqnarray}
as the minimum mass of a stable, strongly interacting, quantum mechanical particle with charge $Q$. Here $M_{\rm sT}$ is the strong gravity analogue of the critical holographic mass scale $M_T = (\hbar^2\sqrt{\Lambda/G})^{1/3}$ investigated in \cite{B4}.

Evaluating this for $Q = \pm (2/3)e$ gives
\begin{eqnarray}  \label{SG:m>me_bound}
M &\gtrsim& (4/9)\alpha_{e}(m_{\rm sP}^2m_{\rm sd})^{1/3} = 2.155 \times 10^{-27} \, {\rm g}
\nonumber\\
&\simeq& m_{\rm u} \simeq 4.457 \times 10^{-27} \, {\rm g} \, ,
\end{eqnarray}
where we have taken the mass of the up quark as $m_{\rm u} \simeq (1.7+3.3)/2 = 2.5$ MeV $= 4.457 \times 10^{-27}$ g.
The numerical estimates in Eqs. (\ref{SG_m_Q}) and (\ref{SG:m>me_bound}) follow directly from the fact that, {\it assuming} $G_f \simeq 10^{38}G$ (\ref{Dirac'sNo.}) and identifying $B \simeq 2 \times 10^{14} \ {\rm gcm^{-3}} \equiv \rho_{\Lambda_f}$ (\ref{B_vs_Lambda_f}), we have $\Lambda_f \simeq 3.351 \times 10^{81}\Lambda$ (\ref{Lambda_f_estimate}), so that
\be  \label{Lambda_f/G_f_vs_Lambda/G}
\frac{\sqrt{\Lambda_f}}{G_f} \simeq 9.154 \times 10^{-21} \ {\rm cm^{-4}gs^{2}} \simeq 5.789 \times 10^2 \frac{\sqrt{\Lambda}}{G} \, .
\ee
Hence, $(\sqrt{\Lambda_f}/G_f)^{1/3} \simeq 8.334 \times (\sqrt{\Lambda}/G)^{1/3}$, so that the estimate for the minimum mass given by Eq. (\ref{SG:m>me_bound}) is only around one order of magnitude higher than the electron mass. According to the strong gravity model -- combined with analogues of the hypothetical MLURs for an asymptotically de Sitter Universe derived in \cite{B2,LakePaterek1,Lake1} -- this should be the minimum possible mass of a stable, strongly interacting, quantum mechanical particle of charge $Q = \pm (2/3)e$.

Reinserting the inequality stemming from Eq. (\ref{SG_Buch_Q_Lambda}) then yields
\begin{eqnarray}  \label{SG:Q^2/mu_bound}
\frac{Q^2}{M} &\lesssim& \left(\frac{3\hbar^2G_f^2c^6}{\Lambda_f}\right)^{1/6} = 3.776 \times 10^{7} \ {\rm Fr^2g^{-1}}
\nonumber\\
&\simeq& \frac{4}{9}\frac{e^2}{m_u} = 2.301 \times 10^{7} \ {\rm Fr^2g^{-1}} \, .
\end{eqnarray}
According to this relation, if the up quark were any less massive (with the same charge $+2e/3$) or more highly charged (with the same mass $m_u$) a combination of canonical quantum pressure and electrostatic repulsion would overcome the strong force attraction (at the outer regions of its mass distribution), destabilising the Compton wavelength. (See analogous arguments made in \cite{B2,LakePaterek1, Lake1} for the canonical gravitational case). Equation (\ref{SG:Q^2/mu_bound}) may also be rewritten as
\begin{eqnarray} \label{Lambda_f_vs_ru}
\Lambda_f \lesssim \frac{l_{\rm sP}^4}{r_u^6} &=& \left(\frac{3}{2}\right)^{12}\frac{3\hbar^2G_f^2m_u^6c^6}{e^{12}}
\nonumber\\
&\simeq& 2.434 \times 10^{26} \ \;{\rm cm}^{-2} \, ,
\end{eqnarray}
where
\be \label{r_u}
r_u = \frac{4}{9}\frac{e^2}{m_uc^2} \simeq 2.558 \times 10^{-14} \;{\rm cm}
\ee
is the {\it classical} radius of the up the quark. Numerically, this is of the order of the size of a fundamental strong gravity `cell' $R_{\rm cell(s)}$, given by Eq. (\ref{SG_R_cell}). Thus, we obtain the analogue of the maximum charge-squared to mass bound in a dark energy Universe, Eq. (\ref{Q^2/me_bound}), for strongly interacting matter, and have demonstrated that the `strong cosmological constant' $\Lambda_f$ may be expressed in a a form analogous to Eq. (\ref{Lambda_vs_re}), i.e. in terms of the relevant `Planck length' and the (classical) radius of a particle that saturates the upper charge-squared to mass ratio bound.

Finally, we note that, in addition to the dimensionless constant $N = (l_{\rm dS}/l_{\rm Pl})^2 = 1.030 \times 10^{122}$, which may be interpreted as the ratio of the the number of `cells' in the three-dimensional bulk space to the number of Planck sized `bits' on the two-dimensional de Sitter boundary, we can construct the strong gravitational analogue,
\be \label{N_f}
N_f 
= \left(\frac{l_{\rm sd}}{l_{\rm sP}}\right)^2 \simeq 307.768 \sim \mathcal{O}(10^2) \, .
\ee
This implies that approximately $10-10^2$ `strong gravity cells', each with linear dimension comparable to the {\it classical} up quark radius, $r_u \simeq R_{\rm cell(s)} \simeq 10^{-14}$ cm, can be packed within the strong de Sitter radius, $l_{\rm sd} \simeq l_{\Lambda_f} \simeq 10^{-13}$ cm. Taking the nucleon radius obtained from scattering cross sections data, $r_n \simeq l_{\rm sP} \simeq \tilde{l}_{\Lambda_f} \simeq 10^{-14}$ cm, this implies that $1-10$ such quarks can exist in a nucleon bound state. This is (obviously) consistent with known physics, but it is interesting to note that such a requirement may also be viewed as a holographic relation for the `de Sitter' horizon of the strong gravity metric. In the holographic picture, the density yielding $\sim 308$ cells per cubic femtometer may be interpreted as the {\it critical density}, above which deconfinement will occur.

In addition, using purely dimensional arguments, we may also define the dimensionless constants
\bea \label{additional_Ns}
N_{\rm Sa} &=& \left(\frac{l_{\rm dS}}{l_{\rm sP}}\right)^2  = \frac{3c^3}{\hbar G_f \Lambda} \simeq 1.031 \times 10^{81} \, ,
\nonumber\\
N'_{\rm Sa} &=& \left(\frac{l_{\rm sd}}{l_{\rm Pl}}\right)^2  = \frac{3c^3}{\hbar G \Lambda_f} \simeq 3.076 \times 10^{44} \, .
\eea
By analogy with Eq. (\ref{Salam_scales}), we christen these the first and second `Salam numbers', respectively, though their physical meaning (if any) remains unclear and further investigations lie beyond the scope of the present work.

Returning again to the Dirac-type relation $G_f \simeq 10^{38}G$ (\ref{Dirac'sNo.}), we now speculate that
\be \label{Dirac'sNo.*}
G_f /G = \alpha_eN^{1/3} \simeq 3.419 \times 10^{38} \, .
\ee
Combining this with Eqs. (\ref{Lambda_f_vs_ru}) and (\ref{B_vs_Lambda_f}), we then have
\bea \label{New_Dirac-1}
B &\simeq& \frac{3}{8\pi}\left(\frac{3}{2}\right)^{12}\frac{\hbar G m_u^6c^7}{e^{10}}\left(\frac{3c^3}{\hbar G\Lambda}\right)^{1/3}
\nonumber\\
&\simeq& 4.463 \times 10^{15} \ {\rm gcm^{-3}} \, .
\eea
where we have used $m_u \simeq 4.457 \times 10^{-27}$g, as before.  Substituting for $\Lambda$ from Eq. (\ref{Lambda_vs_re}) then gives
\bea \label{New_Dirac-2}
B \simeq \frac{3}{8\pi}\left(\frac{3}{2}\right)^{12}\frac{m_u^6c^6}{m_e^2e^6} =2.89 \times 10^{15} \ {\rm gcm^{-3}} \, ,
\eea
where we have used the previous value of $m_u$ together with the standard value of $m_e$. Thus, we have used the strong gravity theory to `derive' two {\it new} large number coincidences, Eqs. (\ref{New_Dirac-1})-(\ref{New_Dirac-2}), linking the physics of strongly interacting particles to the electroweak scale, gravity and dark energy.


\subsection{The Hagedorn temperature for minimum-mass particles, the expanding Universe, and deconfinement}  \label{sect5.3}

As shown in \cite{B1}, $M_{\Lambda}$ may also be interpreted as the effective mass of a dark energy particle. In this picture, the dark energy field is composed of a `sea' of quantum particles, each occupying a volume $V_{\Lambda} \simeq \lambda_{\rm C}^3(m_{\Lambda}) \simeq l_{\Lambda}^3$. Based on this, we now consider an alternative interpretation of the dark energy density and resultant late-time accelerated expansion of the Universe. Though speculative, this interpretation gives rise to a number of interesting results, and it is clear that its analogue in the strong gravity model may be relevant for hadronic physics.

If the dark energy particles are charge neutral and are their own antiparticles, then, under these conditions (in which the average inter-particle distance is comparable to the Compton wavelength $\sim \hbar/(M_{\Lambda}c)$), standard quantum theory implies that they will readily pair-produce. However, this is impossible without a concomitant expansion of space itself. In this picture, otherwise `empty' space is full of dark energy particles, which give rise to an effective constant energy density on large scales. Borrowing a term from basic chemistry to describe this state, we may say that the space is `saturated', and remains so as more space is produced to `absorb' the newly created particles. Thus, the creation of dark energy particles and of space-like quanta go hand in hand.

It is straightforward to see that, if the probability of pair-production remains constant, the scale factor of the Universe $a(t)$ will grow exponentially, since the number of particles produced by a given volume, per unit time, is proportional to the volume itself. Let us assume that, together with the pair-production of a single dark energy particle, $n_{\rm cell}$ new fundamental `cells' of space are also produced, with total volume $V =n_{\rm cell}V_{\rm cell} \simeq V_{\Lambda}$. In \cite{B2}, it was already shown that, if there exists a holographic relation between the bulk and the boundary of our (asymptotically) de Sitter Universe, such fundamental cells must have linear dimension of order
\bea \label{R_cell}
R_{\rm cell} &\simeq& (l_{\rm Pl}^2l_{\rm dS})^{1/3} \simeq 3.500 \times 10^{-13} \ {\rm cm}
\nonumber\\
&\simeq& r_e = e^2/(m_e c^2) = 2.818 \times 10^{-13} \ {\rm cm} \, .
\eea
This is also the Compton radius associated with the critical mass scale $M_T=(\hbar^2\sqrt{\Lambda}/G)^{1/3}$, investigated in \cite{B4}, and the Schwarzschild radius of the dual mass, $M_T^{'} = m_{\rm Pl}^2/M_T = c(\hbar /G^2\sqrt{\Lambda})^{1/3}$. Equation (\ref{R_cell}) ensures that the number of cells in the bulk is equal to the number of Planck sized bits on the de Sitter boundary, i.e.
\be \label{N}
N = \frac{V_{\rm dS}}{V_{\rm cell}} = \left(\frac{l_{\rm dS}}{l_{\rm Pl}}\right)^2 \simeq \frac{3c^3}{\hbar G \Lambda} \simeq 1.030 \times 10^{122} \, .
\ee

Next, let us suppose that the probability of a single cell of space `pair-producing' within a time $t_{\rm Pl} = l_{\rm Pl}/c$, due to the presence of the dark energy density, is given by
\bea
&&P(\Delta V = +V_{\rm cell}|V_0=V_{\rm cell},\Delta t =t_{\rm Pl}) = N^{-1/2}
\nonumber\\
&& = \frac{l_{\rm Pl}}{l_{\rm dS}} = \frac{V_{\rm Pl}}{V_{\rm cell}}  \simeq \left(\frac{\hbar G \Lambda}{3c^3}\right)^{1/2} \simeq 9.851 \times 10^{-62} \, .\notag \\
\eea
This leads naturally to a de Sitter-type expansion, modelled by the differential equation
\bea \label{dS_exp-1}
&&\frac{d a^3}{dt} = \frac{N^{-1/2} a^3}{t_{\rm Pl}} = \frac{l_{\rm Pl}}{l_{\rm dS}}\frac{a^3}{t_{\rm Pl}}
\nonumber,
\eea
or, equivalently,
\bea \label{dS_exp-2}
\frac{d a}{dt} \simeq c\sqrt{\frac{\Lambda}{3}}a \, ,  \quad a(t) \simeq a_0e^{-c\sqrt{\Lambda/3}t} \, .
\eea
In this picture, the macroscopic dark energy energy density $\rho_{\Lambda}$ remains approximately constant, in spite of spatial expansion, the additional (positive) mass-energy of a newly created dark energy particle being exactly counterbalanced by the additional (negative) energy contained in its gravitational field. This may be shown explicitly by considering the Komar energy (see \cite{B1,B2,B3}). 

However, if this picture is correct, we may expect `empty' three-dimensional space to exhibit granularity on scales $\sim l_{\Lambda}$. It is therefore particularly intriguing that recent experiments provide tentative hints of fluctuations in the gravitational field strength on scales comparable to $l_{\Lambda} = \hbar/(M_{\Lambda}c)$, which is of order $0.1$ mm \cite{Perivolaropoulos:2016ucs,Antoniou:2017mhs}. Though many theoretical models may account for this, including those exhibiting spatial variation of Newton's constant $G$, the results presented above imply that the `granularity' of the dark energy density, due to the presence of effective dark energy particles on sub-millimetre scales, cannot be discounted {\it a priori.}

In this model, the number of holographic spatial cells created when one dark energy particle is pair-produced is $n_{\rm cell} \simeq (\hbar G \Lambda/3c^3)^{1/4} = N^{1/4} \simeq 3.186 \times 10^{30}$. As already noted in \cite{B1}, this number is also the multiplying factor that naturally generates a sequence of mass scales between $m_{\rm dS}$ and $m'_{\rm dS}$, i.e. $m'_{\rm dS} = N^{1/4}M'_{\Lambda} = N^{1/2}m_{\rm Pl} = N^{3/4}M_{\Lambda} = Nm_{\rm dS}$, where $M'_{\Lambda} \equiv m_{\rm Pl}^2/M_{\Lambda}$.

In addition, it is clear that the fundamental field giving rise to the dark energy density (whatever its precise nature may be) remains `trapped' in a Hagedorn phase.  Any attempt to further compress (i.e. heat) the `sea' of dark energy particles -- even if such compression results simply from random quantum fluctuations -- results in pair-production rather than increased kinetic energy. The saturation condition implies the existence of not-so-UV cut-off for the vacuum field modes, given by $\lambda_{\rm DE} \simeq \lambda_{\rm C}(M_{\Lambda}) = l_{\Lambda}$, yielding
\bea \label{rho_vac}
\rho_{\rm vac} \simeq \frac{\hbar}{c} \int_{1/l_{\rm dS}}^{1/l_{\Lambda}} \sqrt{k^2 + \left(\frac{2\pi}{l_{\Lambda}}\right)^2} d^3k
\nonumber\\
\simeq \frac{m_{\rm Pl}l_{\rm Pl}}{l_{\Lambda}^{4}} \simeq \frac{\Lambda c^2}{G} \simeq 10^{-30} \ {\rm gcm^{-3}} \, .
\eea
Thus, the temperature associated with the field remains constant, on large scales, and is comparable to the present day temperature of the CMB \cite{B1},
\be \label{T_Lambda}
T_{\Lambda} \equiv \frac{M_{\Lambda}c^2}{8 \pi k_{\rm B}} \simeq 2.27 \ {\rm K} \simeq T_{\rm CMB} = 2.73 \ {\rm K} \, .
\ee
Here the factor of $(8\pi)^{-1}$ is included in the definition of $T_{\Lambda}$ by analogy with the standard expression for the Hawking temperature, yielding $T_{\Lambda}(M_{\Lambda}) = T_{\rm H}(M'_{\Lambda})$, where $M'_{\Lambda} = m_{\rm Pl}^2/M_{\Lambda}$ again denotes the dual mass.

Though this too may seem like another incredible coincidence, we note that, in the dark energy model implied by the DE-UP (\ref{DE-UP}), it is simply a restatement of the standard �coincidence problem� of cosmology, i.e. the Universe begins a phase of accelerated expansion when $r_{\rm U} \simeq l_{\rm dS}$, at which point $\Omega_{\rm M} \simeq \Omega_{\Lambda}$ and, hence, $T_{\rm CMB} \simeq T_{\Lambda}$. The question remains, why do we live at precisely this epoch? However, no {\it new} coincidences are required, in order to explain Eq. (\ref{T_Lambda}).


The implications of this picture for the dual strong gravity model are self-evident. In this case, the mass associated with the dark energy `cell' $M_{\Lambda}$ is replaced by the nucleon mass $m_n \simeq M_{\Lambda_f} \simeq \tilde{M}_{\Lambda_f}$. When compressed beyond the nuclear density, a free quark fluid is formed, and further attempts at compression (i.e. heating) simply result in the production of more strongly interacting matter (quark-gluon plasma). The free strange quark matter remains `locked' in a Hagedorn phase and the temperature of the plasma remains constant, given by
\bea \label{T_Lambda}
T_{\Lambda_f} &\equiv& \frac{\tilde{M}_{\Lambda_f}c^2}{8\pi k_{\rm B}} \simeq \frac{\tilde{M}_{\Lambda_f}}{M_{\Lambda}}T_{\Lambda}
\nonumber\\
&\simeq& 10^{12}T_{\Lambda} \simeq 10^{12} \ {\rm K} \, .
\eea
This is consistent with previous estimates for the temperature of the deconfinement transition obtained in Sec. \ref{sect4}.

\section{Discussions and final remarks}\label{sect6}

In the present paper, we have considered the mass/radius ratio bounds for spherical compact objects with anisotropic pressure distributions in the strong gravity model, which represents an attempt to describe the gauge singlet sector of the strong interaction using a `geometric' theory, based on analogy with general relativity. Though the theory cannot describe the $SU(3)$ colour charge sector of QCD, it remains a viable candidate for an effective theory which may be used to model stability conditions and confinement in strongly interacting particles.

Strong gravity is a two tensor theory, in which the canonical (weak) gravitational field is described by the usual spacetime metric $g_{\mu\nu}$ and the strong interaction is described by an additional metric-type tensor $f_{\mu\nu}$. The strong gravity action contains the usual Einstein-Hilbert term $(1/2k_f^2)R(g)\sqrt{-g}$, where $k_g^2 = 8\pi G/c^4$ plus an additional `copy' constructed from the strong tensor field, $(1/2k_f^2)R(f)\sqrt{-f}$, where $k_f^2 = 8\pi G_f/c^4$ and $G_f \simeq 10^{38}G$ is the strong gravity coupling constant. The strong gravity Lagrangian also includes an interaction term $\mathcal{L}_{fg}$ and the standard matter term $\mathcal{L}_{m}$. The interaction term is proportional to $\mathcal{M}^2$, where $\mathcal{M}$ is the mass of the spin-2 `strong graviton'.

An appropriate choice of `gauge' for the mixing term generates an effective {\it strong gravitational constant}, $\Lambda_f \propto \mathcal{M}^2$. Hence, in the strong gravity theory, there exist analogues of many results that can be derived from canonical general relativity with a `true' cosmological constant term, $\Lambda$. These include mass bounds obtained using the appropriate Buchdahl-type inequalities for the physical system under consideration. Therefore, in strong gravity, it is possible to obtain explicit inequalities giving upper and lower bounds on the ratio $M_{{\rm eff}}/R$, where $M_{\rm eff}$ is the effective mass of a compact object, including the contribution from $\Lambda_f$. Alternatively, these may be reformulated as bounds on $M_{0}/R$, where $M_0$ is the bare mass, and the inequality is written as an explicit function of the strong cosmological constant and the (pressure) anisotropy parameter $\mathcal{D}$, which also depends on $\Lambda_f$.

As is the case for compact objects in general relativity, in the presence of a nonzero cosmological constant ($\Lambda \neq 0$), we found two universal limits (upper and lower) for the mass/radius ratio of strongly interacting particles. However, due to the presence of the strong cosmological constant ($\Lambda_f \neq 0$) and of the anisotropies in the pressure distribution, the physical and geometric properties of such hadronic-type compact objects are significantly modified within the particle interior, as compared to their counterparts in standard general relativity. Both the upper and lower mass/radius ratios depend sensitively on the values of $\Lambda_f $ and $\mathcal{D}$ at the surface of the hadron.

In addition, different physical models for the mixing term that generates the effective strong cosmological constant can lead to very different mass/radius relations. It is a general feature of the behaviour of the physical and geometrical parameters of anisotropic objects in strong gravity that the increase in mass is proportional to the deviations from isotropy, described in our approach by the function $\mathcal{D}$. Since these deviations from isotropy are arbitrary, there are no mathematical or theoretical restrictions that restrict the radii of hadronic-type strong gravity structures, which may therefore extend up to the `apparent horizon' of the strong tensor field. Under the assumption that the exterior of the hadronic objects is described by the strong Schwarzschild metric, this corresponds (approximately) to the mass $M_{{\rm eff}} \leq c^2R/(2G_f)$.

Usually, the upper mass/radius bound can be obtained when we assume the ultra-stiff equation of state, i.e. constant density inside the sphere.  For the minimum mass/radius bound, a straightforward interpretation~(in the de Sitter case) is the situation where we have constant pressure equal to the vacuum pressure throughout the sphere so that the pressure is always balanced inside and outside.  Constant pressure implies $\rho=-P$ inside the sphere and this is simply the vacuum density. The minimum mass/radius ratio actually implies the minimum density of the object, which is equal to the vacuum density due to the presence of dark energy. It actually implies that no particle should have smaller density than the vacuum density.  

The Schwarzschild-de Sitter-type solution of the strong gravity field equations describes a microscopic system embedded in an ordinary, flat space-time, in which the mass of compact coloured objects is localized due to the `curvature' of the strong metric field, which creates a kind of `bag' \cite{sg3}. By interpreting the energy density of the strong gravity cosmological constant as the bag constant of QCD \cite{Shur}, it follows that strong gravity imposes the following {\it classical} lower bound on the minimum mass/radius ratio of a hadron,
\be \label{genBuch}
\frac{2G_fM_0}{c^2R}\geq \frac{1}{6}\Lambda_fR^2 \, .
\ee
This is related to the bag constant $B$, via $B \simeq \rho_{\Lambda_f} \simeq \Lambda_fc^2/(8\pi G_f)$ (\ref{B_vs_Lambda_f}). By assuming a hadronic radius of the order of $R=0.8$ fm (comparable to the proton radius \cite{Pohl}), and taking the estimate $\Lambda_f \simeq 10^{25}$ cm$^{-2}$, obtained from the identification (\ref{B_vs_Lambda_f}) together with Eq. (\ref{Dirac'sNo.}), we obtain a lower bound on the bare mass of a hadron as $M_0 \gtrsim 10^{-24}$ g, a value which is of the same order of magnitude as the mass of a nucleon. Smaller particle radii, of the order of $0.1$ fm, will give considerable lower hadron masses, of the order of $M_0\geq 3.37\times 10^{-28}$ g.

All results regarding the mass/radius ratios for anisotropic hadronic objects have been obtained by assuming the basic conditions (\ref{condinq}) and (\ref{46}). However, for an arbitrarily large anisotropy parameter $\mathcal{D}$, with $P_r \gg P_{\perp}$, we can not exclude (in principle) the situation in which these conditions do not hold. If, for example
\begin{equation}
\gamma \left(r'\right)<\gamma (r),\forall r>r' \, ,
\end{equation}
then, for a hadron in strong gravity with monotonically decreasing density, the condition
\begin{equation}
\frac{d^{2}\Psi }{d\xi ^{2}}>0, \quad \forall r \, .
\end{equation}
holds in place of Eq. (\ref{46}). This situation corresponds to a tangential pressure-dominated hadronic  structure, with the tangential increasing inside the compact object. In this case, we obtain a restriction on the minimum mass/radius ratio so that, for this hypothetical, ultra-compact hadronic particle, $4/9$ is an {\it absolute} lower bound for the value of $G_fM/(c^2R)$.

In addition, we have investigated possible quantum mechanical implications of the strong gravity model, for both neutral and charged particles. For neutral particles, the quantum minimum mass bound follows by identifying the classical radius $R$ in Eq. (\ref{genBuch}) with the Compton wavelength $\lambda_{\rm C}$. The mass scale thus obtained is roughly comparable to the mass of the neutron, $m_{n} \simeq 10^{-24}$g, the lightest known stable, compact, charge-neutral and strongly interacting particle.

To treat charged hadronic objects, we combined classical stability bounds for charged compact objects in strong gravity, obtained by substituting $G \rightarrow G_f$ and $\Lambda \rightarrow \Lambda_f$ into their general-relativistic counterparts, with hypothetical minimum length uncertainty relations (MLURs). These, in turn, were based on MLURs obtained by considering canonical gravitational `corrections' to the standard Heisenberg uncertainty principle, including the effects of dark energy and the existence of a de Sitter horizon \cite{B2,LakePaterek1,Lake1}. The formal similarity between general relativity and the strong gravity theory again allowed us to replace $G \rightarrow G_f$ and $\Lambda \rightarrow \Lambda_f$, yielding analogous MLURs for strongly interacting particles. Identifying $R \simeq (\Delta x_{\rm total})_{\rm min}$, and evaluating the strong gravity MLUR Eq. (\ref{SG_DE-UP}) for $Q=\pm 2e/3$, we obtained an estimate of the mass of the up quark, $m_u \simeq 10^{-27}$g, the lightest known stable, compact, charged and strongly interacting particle. This estimate is {\it equivalent} to a new large number coincidence, `derived' from the quantum mechanical MLUR/strong gravity model, which relates the nuclear density, dark energy density, and physics at the electroweak scale, Eq. (\ref{New_Dirac-1}).

The formal equivalence between the mathematical structure of the strong gravity theory and canonical general relativity also permits us to draw parallels between the strong gravity model of quark deconfinement and the expansion of the Universe in the particle `sea' model of dark energy, proposed in \cite{B1}. In the former, an expanding deconfined quark matter remains `trapped' in a Hagedorn phase, in which further compression of the quark-gluon plasma, {\it even} if this arises as a result of random quantum fluctuations, leads to pair-production rather than increased temperature. By interpreting the minimum mass of a stable, compact, charge-neutral, and quantum mechanical object as the mass of an effective dark energy particle \cite{B1}, we obtained the resulting `Hagedorn temperature' ($T_{\Lambda}$) of the dark energy field, which was found to be comparable to the present day temperature of the CMB (\ref{T_Lambda}). In this model, such a `coincidence' is not really a {\it new} coincidence at all, but simply a restatement of the standard coincidence problem in cosmology, whereby a phase of accelerated expansion begins when $r_{\rm U} \simeq 1/{\sqrt{\Lambda}}$ and $\Omega_{\rm M} \simeq \Omega_{\Lambda}$, or, equivalently, $T_{\rm CMB} \simeq T_{\Lambda}$.

Finally,  we note again that, since there is no explicit SU(3) gauge symmetry in the strong gravity field equations, these may describe only the gauge singlet sector of the strong interaction, mediated by massless and massive spin-2 particles, coupled to the energy-momentum tensor of the strongly interacting matter. Hence, strong gravity is not expected to replace QCD, but may be used to describe interactions involving only gauge singlet states, using a gravitational type formalism, though {\it not} the sector including colour charges. It is therefore justified to use strong gravity to explore the stability and confinement of gauge singlet mesons and baryons, but not scattering processes that require colour charge interactions.

\acknowledgments

We would like to thank to the anonymous referee for comments and suggestions that helped us to improve our manuscript.  P.B. is supported in part by the Thailand Research Fund~(TRF) and Chulalongkorn University under grant RSA5780002. M. L. was supported by a Naresuan University Research Fund Individual Research Grant during the preparation of this work and by the gracious hospitality of Sun Yat-Sen University and Nanyang Technological University.


\end{document}